# All-optical method to directly measure the pressure-volume-temperature equation of state of fluids in the diamond anvil cell

J.E. Proctor*, C.E.A. Robertson, L.J. Jones, J. Phillips, K. Watson, Y. Dabburi and B. Moss

Materials and Physics Research Group, University of Salford, Manchester M5 4WT, United Kingdom

* Corresponding author: j.e.proctor@salford.ac.uk

**Abstract**

We have developed a new all-optical method to directly measure the pressure-volume-temperature (PVT) equation of state (EOS) of fluids and transparent solids in the diamond anvil high pressure cell by measuring the volume of the sample chamber. Our method combines confocal microscopy and white light interference with a new analysis method which exploits the mutual dependence of sample density and refractive index: Experimentally, the refractive index determines the measured sample chamber thickness (and therefore the measured sample volume/density), yet the sample density is by far the dominant factor in determining the variation in refractive index with pressure. Our analysis method allows us to obtain a set of values for the density and refractive index which are mutually consistent, and agree with the experimental data within error. We have conducted proof-of-concept experiments on a variety of samples ($H_2O$, $CH_4$, $C_2H_6$, $C_3H_8$, KCl and NaCl) at ambient temperature, and at high temperatures up to just above 500 K. Our proof-of-concept data demonstrate that our method is able to reproduce known fluid and solid EOS within error. Furthermore, we demonstrate that our method allows us to directly and routinely measure the PVT EOS of simple fluids at GPa pressures up to, at least, 514 K (the highest temperature reached in our study). A reasonable estimation of the known sources of error in our volume determinations indicates that the error is currently ± 2.7% at high temperature, and that it is feasible to reduce it to ca. ± 1% in future work.

1. Introduction

The pressure-volume-temperature (PVT) equation of state (EOS) of a substance is a material property of fundamental importance. However, there is a basic shortcoming in humankind's ability to measure this property for fluids. Direct EOS measurements in real space using piston-cylinder devices have been made on most simple fluids (e.g. $CH_4$ [1], $CO_2$ [2], Ar [3], $N_2$ [4]) up to the fusion (liquid-solid transition) (typically 1 – 2 GPa) at ca. 300 K. However, with increasing temperature fusion pressures increase rapidly while the pressures achievable in piston-cylinder devices decrease rapidly (see, for example the measurements on $CH_4$ in ref. [1]). Pressures adequate to freeze simple fluids above 300 K can only be generated using opposed anvil devices (Paris-Edinburgh cell, Multi-anvil press and – principally – the diamond anvil cell (DAC)). In these devices direct volume measurement in real space is not possible. The PVT EOS of solids contained in these devices can be determined by measurements in reciprocal space using X-ray or neutron diffraction to directly determine the lattice constants and structure. However, fluid PV EOS cannot be determined using this method either at 300 K or at high temperature due to the lack of Bragg peaks from fluids.

There is therefore a massive gap in humankind's knowledge of the PVT EOS of simple fluids. For instance, in $CH_4$ at 500 K directly measured PV EOS data are available up to 40 MPa, yet the freezing pressure is ca. 3500 MPa ([5] and refs. therein). Various approaches have been attempted over the decades to rectify this, yet none have been widely adopted and it seems that all have serious shortcomings.

The first approach is simply to extrapolate from the low pressure data. Commonly used software packages such as NIST REFPROP / WEBBook [6] routinely provide PVT EOS output that are in fact extrapolations from the real experimental data collected at lower pressures and/or lower temperatures. For instance, for $CH_4$ PVT EOS output is provided up to 1000 MPa at 625 K when the real experimental data at 625 K only extend to 40 MPa and the real experimental data to 1000 MPa exist only at 300 K [1].

These extrapolations are performed using a model called the "Fundamental Equation of State" (described in ref. [5] and citations therein) that incorporates ca. 50 dimensionless and physically meaningless adjustable parameters and even has a mathematical form that is partially empirical. The "fundamental equation of state" model was designed to be overfitted to allow accurate interpolation between real datapoints so using it to extrapolate is clearly not ideal.

The second approach is Brillouin spectroscopy in the diamond anvil cell. In this approach the speed of sound $v$ in the fluid is measured. Usually the bulk modulus is obtained from this ($B = \rho v^2$ where $B$ is the bulk modulus and $\rho$ is the density), which is then integrated to obtain the $PV$ EOS along the isothermal paths typically followed in an experiment (see, for instance, ref. [7]). However, sound waves propagate adiabatically so the parameter that directly follows from the speed of sound is the adiabatic bulk modulus. To convert between the adiabatic and isothermal bulk moduli it is necessary to know the heat capacity $c_P$ and thermal expansion coefficient $\alpha$. Fluid Brillouin spectroscopy EOS studies deal with this problem by integrating the adiabatic bulk modulus whilst neglecting the adiabatic → isothermal correction then obtaining the heat capacity and thermal expansion coefficient values from the result of this integration, using the thermodynamic relations shown below in equation (1). The integration is performed incorporating the adiabatic → isothermal correction (the $T\alpha^2/c_P$ term in equation (1)) using these values and repeated iteratively until convergence is achieved. This iterative process is repeated for each gap between datapoints in the experiment. Thus no direct measurement of volume versus pressure is made, and a single error will affect all subsequent pressure points in the experiment.

$$\left(\frac{\partial \rho}{\partial P}\right)_T = \frac{1}{v^2} + \frac{T\alpha^2}{c_P}$$

$$\alpha = -\frac{1}{\rho}\left(\frac{\partial \rho}{\partial T}\right)_P$$

$$\left(\frac{\partial c_P}{\partial P}\right)_T = -\frac{T}{\rho}\left[\alpha^2 + \left(\frac{\partial \alpha}{\partial T}\right)_P\right]$$

(1)

In some cases, this problem is avoided by calculating the density via the sample refractive index (RI) $n$ instead [8][9]. The RI can be obtained using Brillouin spectroscopy if data are collected for different scattering geometries, and the density can be calculated from the RI using the Lorentz-Lorenz law [10] (shown in equations (2) and (5)). However, this just introduces a different source of error: The assumption that the value of the Lorentz-Lorenz factor $L$ which provides the best fit to the $n(\rho)$ data at lower pressure and temperature continues to be the best-fit value over the wider PT range covered by the Brillouin spectroscopy experiment. In at least one case, this assumption has been shown to produce an EOS [9] that cannot possibly be correct since it would lead to a violation of the Clausius-Clapeyron equation at the solid-liquid (fusion) phase transition [11]. The Lorentz-Lorenz law is written in the following form to determine the Lorentz-Lorenz factor $L$ from experimental RI measurements at lower pressures, where the density is known:

$$L\rho = \frac{n^2 - 1}{n^2 + 2}$$

(2)

The density can then be determined from the RI at higher pressures. It is clear from equation (2), and has been noted previously [12], that when utilizing RI data collected in the gas phase to determine $L$ a small error in the RI will lead to a large error in $L$ since the RI is close to 1.

Furthermore, validating the experimental and analysis methods used to produce EOS using Brillouin spectroscopy is challenging due to the difficulty of conducting appropriate control experiments on substances with a known EOS. Direct EOS measurements using piston-cylinder apparatus are usually only available to pressures so low that they are hard to control and measure in the DAC. Control experiments reproducing known transparent solid EOS are of limited utility since the adiabatic → isothermal correction to the bulk modulus (a weak link in the analysis method for Brillouin spectroscopy data) is negligible for solids but is certainly not negligible for fluids. Instead, Brillouin spectroscopy EOS are commonly compared to simulations, extrapolations and other Brillouin spectroscopy studies.

The third approach is to determine the EOS by making direct measurements of the volume of the entire sample chamber at different pressures [12][13][14]. The DAC sample chamber in a fluid EOS experiment is roughly cylindrical, with a thickness $t$ (i.e. distance between the diamond culets) of ca. 50 - 100 μm and a diameter of ca. 200 – 400 μm so this measurement can be performed using imaging and interference phenomena involving visible light. The cross-section can be determined by recording a microscope image of the sample chamber with transmitted light only and counting the bright pixels using image processing software. The first stage of the process to determine $t$ is to measure the product $nt$. To do this, a spectrum is collected of white light transmitted through the sample chamber along the axis perpendicular to the diamond culets. The spectrum consists of interference fringes corresponding to wavelengths at which light that has been scattered back and forth across the sample chamber interferes constructively or destructively with light that has passed straight through. The product $nt$ (where $n$ is the sample RI) can be calculated trivially by measuring the wavelength spacing between the fringes. We conducted some preliminary experiments (shown in the supplementary material) on NaCl at 295 K to 8 GPa. The RI of NaCl at high pressure is known [15], and we found that – using the known RI – obtaining a $PV$ EOS that agrees with that determined from X-ray diffraction [16] from the white light interference fringes and cross-section images was straightforward. Clearly, determining the RI is the most difficult aspect of this approach to determine fluid EOS.

Three methods have been developed to obtain the RI. The first [12][13] is to exploit the parallelism of the diamond culets by observing Fabry-Perot interference rings when the sample chamber is illuminated with diffuse monochromatic light. In this approach the angles at which the rings are observed are determined in part by refraction at the sample – diamond and diamond - air boundaries. The influence of the RI via Snell's law therefore allows it to be determined independently of $t$. The second [14][17][18] is to measure the reflectivity of the diamond – sample boundary which is determined by the RIs of the relevant media according to Fresnel's law [19]. The third [17][18][20] is to illuminate the DAC with a focussed laser beam along the axis perpendicular to the culets and to measure the distance moved by the DAC between the points where the beam is focussed on the piston and cylinder culets (this distance is not identical to $t$ since refraction at the diamond-sample boundary must be accounted for, but the RI and $t$ can be obtained by combining this measurement with one of the other measurements listed above). This method is referred to as confocal microscopy.

To our knowledge, fluid EOS measurements based on determination of the DAC sample chamber volume have been performed only at ambient temperature. In a small number of studies, the RI has been determined at high temperature using the Fabry-Perot fringes method [7][21], but the data were not combined with measurements of the sample chamber cross-section to generate an EOS. Instead the RI data were used to allow sound speed calculation from the 180° backscattering geometry in Brillouin scattering, rather than the platelet geometry (in the platelet / forward scattering geometry, the sound speed can be measured without knowing the RI of the sample [22], but the diamond culet parallelism requirements are probably more stringent). The confocal microscopy method has been combined with the Fresnel method to produce EOS of Ar [18] and $H_2O$ [17] at ambient temperature. It has been combined with the white light interference measurement of $nt$ to measure the RI of $H_2O$ at a single pressure (0.3 GPa) [20]. To our knowledge, it has not been attempted at high temperature.

The Brillouin spectroscopy approach has been attempted at high temperature but (even if one takes the complex analysis method at face value) there are few studies in the literature considering the large amount of work to be done in this field, and the importance of fluid EOS experiment and theory to geoscience, planetary science and fundamental physics.

We therefore set out to conduct fluid EOS measurements at high temperature by measuring the volume of the entire sample chamber. Initially we attempted to do this by replicating the Fabry-Perot fringes method. However, we discovered that the requirement for the parallelism of the diamond culets to observe the (monochromatic) Fabry-Perot fringes is very stringent. We attempted to observe these fringes using a number of different DACs in our laboratory, and discovered that even DACs that exhibit excellent quality white light fringes exhibit very poor quality Fabry-Perot fringes. To obtain good quality Fabry-Perot fringes it was necessary to build a DAC in which it was possible to perform tilt alignment of one of the diamonds, following the design principles in ref. [23]. However, to achieve this it was necessary to resort to an even tighter piston-cylinder alignment than usual, and the use of additional components characterized by intricate and delicate kinetic mechanisms. Based on our experience of high temperature experiments in the DAC, we feel that combining this level of intricacy and tight-fitting moving parts with resistive heating is not a very practical solution to the problems outlined above. The requirement to heat the entire sample chamber, of course, precludes the use of laser heating.

It was therefore necessary to switch to a different combination of experimental methods. We chose to combine the white light interference method for the determination of $nt$ with the confocal microscopy method. To our knowledge this combination has only been utilized to measure the RI of water at a single $P, T$ point (0.3 GPa, ambient temperature) [20], and has not yet been utilized to measure an EOS.

In our proof-of-concept experiments we compare EOS determined using this combination of methods to fluid EOS determined using piston-cylinder devices and solid EOS determined using X-ray diffraction. We also conduct experiments at high temperature up to 514 K, demonstrating that our method can be used to determine PVT EOS at least up to this temperature. In addition, we introduce a new analysis method that makes use of the mutual dependency between the sample density and RI: Experimentally, the measured value of the RI determines the measured value of the density, but on the other hand the density is by far the dominant factor in determining the RI via the Lorentz-Lorenz law. Our analysis method exploits this mutual dependence, enabling us to produce a set of values for the density and RI which are mutually consistent, and within a reasonable estimate for the margin of error in our experimental data. Finally, we have quantified the different sources of error in our experiments to indicate how the overall error can be reduced, marking a clear way forward to highly accurate fluid EOS determinations in the future.

## 2. Methods

High pressure and high temperature were generated using DACs equipped with external resistive heaters. Optical spectra were collected using a grating spectrometer and sample cross-sections ($A$) were determined by counting the bright pixels using the open-source Gnu image processing software. Pressure was measured using photoluminescence from a small ruby crystal inside the sample chamber at ambient temperature [24], and using photoluminescence from a small Samarium-doped Strontium Borate (Sm:SrB$_4$O$_7$) crystal in the sample chamber at high temperature [25]. H$_2$O reacts with Sm:SrB$_4$O$_7$ at high temperature combined with high pressure [26], so for our H$_2$O experiments we measured pressure above 1 GPa using a small crystal of Sm:SrB$_4$O$_7$ placed between the diamond and gasket close to the sample chamber. In our second experiment on H$_2$O at 500 K we increased pressure to beyond the freezing point and visually observed freezing at the correct pressure within experimental error. Further experimental details are given in the supplementary material.

The product $nt$ (henceforth labelled as $\delta$) was measured by interference fringes (in the wavelength / wavenumber domain) from white light transmitted through the DAC directly along the optical axis (see experimental geometry in figure 1). We typically counted over 10 – 40 fringes and fitted Gaussian peaks to determine the wavelength of the highest ($\lambda_A$) and lowest ($\lambda_B$) wavelength fringes in the selection, after subtraction of a linear baseline in each case. We experimented with other fitting methods and found that the choice of fitting method has a negligible effect on the obtained value of $\delta$. Equation 3 (derived in the supplementary material) is used to obtain $\delta$ from $\lambda_{A,B}$. Here, $p$ is the number of fringes.

$$\delta = nt = \frac{(p-1)\lambda_A \lambda_B}{2(\lambda_B - \lambda_A)}$$

(3)

The confocal microscopy method, allowing us to determine $n$ (the RI) and $t$ (the sample thickness) separately, exploits the fact $t$ is typically 50 – 100 μm, large compared to all dimensions of a visible laser focal spot size. We shine a collimated red laser beam into the DAC via an objective lens and position the DAC to observe (using a CCD camera) when the red laser beam is focussed on the culet of the cylinder diamond. We then move the DAC so that the beam is focussed on the culet of the piston diamond. The sample stage is connected to a digital micrometer, allowing us to record the distance moved by the DAC. This distance (henceforth labelled as $t'$) is not identical to the thickness $t$, but it can be related to $t$ by accounting for refraction via Snell's law at the diamond-sample and air-diamond interfaces. In the paraxial approximation, equation (4) allows $n$, $t$ and the sample chamber volume $V_S$ to be obtained from our experimental measurements.

$$t = \sqrt{\delta t'}$$

$$n = \sqrt{\frac{\delta}{t'}}$$

$$V_S = At$$

(4)

These equations are derived in the supplementary material, along with equations for the errors in $n$, $t$ and $V_S$ that result from the raw errors in our experimental measurements. Equation (4) can be rewritten as $t' = t/n$. The equation in the $t' = t/n$ form was derived using a different methodology in 2009 by Hanna and McCluskey [17].

To derive equation (4), we have chosen to work in the paraxial approximation (our objective lens has a numerical aperture (NA) of 0.30). The insertion of any plane surface (such as the air-diamond interface) into the beam between the objective lens and the focal point causes the focal point to become smeared out along the optical axis (longitudinal aberration) if non-paraxial effects are significant [27]. We estimate that, for an NA of 0.30 and typical values of $n$ and $t'$, neglecting the effect of longitudinal aberration could cause a systematic error in $t'$ of 0.2 µm (see calculation in supplementary material). Due to the extremely small size of this error we have neglected longitudinal aberration throughout. On the other hand, working in the paraxial approximation can potentially lead to significant random errors since the effect we rely on to determine the RI (the change in angle due to Snell's law at the diamond-sample interface) is by definition a small effect. Working outside of the paraxial approximation (i.e. exploiting the full ± 45° opening angle / NA of 0.7 that is possible using Boehler-Almax diamond seats) would be possible as an alternative method. However our calculations clearly show that the systematic error caused by neglect of longitudinal aberration would then become significant (above 1 µm). This systematic error would be hard to precisely calculate as it would require knowledge of the laser beam profile. On balance, it would therefore seem better to work in the paraxial approximation utilizing a lens with a small NA to reduce the longitudinal aberration.

In some preliminary experiments, we collected data on pressure decrease. However, it was clear from the trends in the sample chamber volume (calculated using equation (4)) versus pressure that sample leakage was occurring from time to time upon pressure decrease, even in the solid state. We therefore strongly advise against utilising volume measurements performed upon pressure decrease, even in the solid state. RI measurements, could in principle be performed upon pressure decrease, however there seems to be more scatter in RI measurements collected on pressure decrease. All measurements of RI and volume presented here were therefore performed on pressure increase. The exceptions to this were some preliminary experiments on $C_2H_6$ at ca. 380 K and $C_3H_8$ at 295 K in which we measured the RI only, and on NaCl at 295 K in which we measured the EOS using the known RI, combining data collection on pressure increase and decrease in all three cases. These data are shown in the supplementary material.

Figure 1 shows illustrative examples of the measurements of $\delta$ and $t'$. As written in equation (4), $V_S$ is in arbitrary units since we do not know how many moles of sample are in the chamber. We will deal with procedures for converting it to absolute units later on a case-by-case basis for each of our datasets. For future determinations of fluid EOS, the best approach is probably to begin at low pressure where the EOS is known, and end by collecting data in the solid state. If the solid state EOS is also known, this provides two independent calibrations for the amount of sample in the chamber.

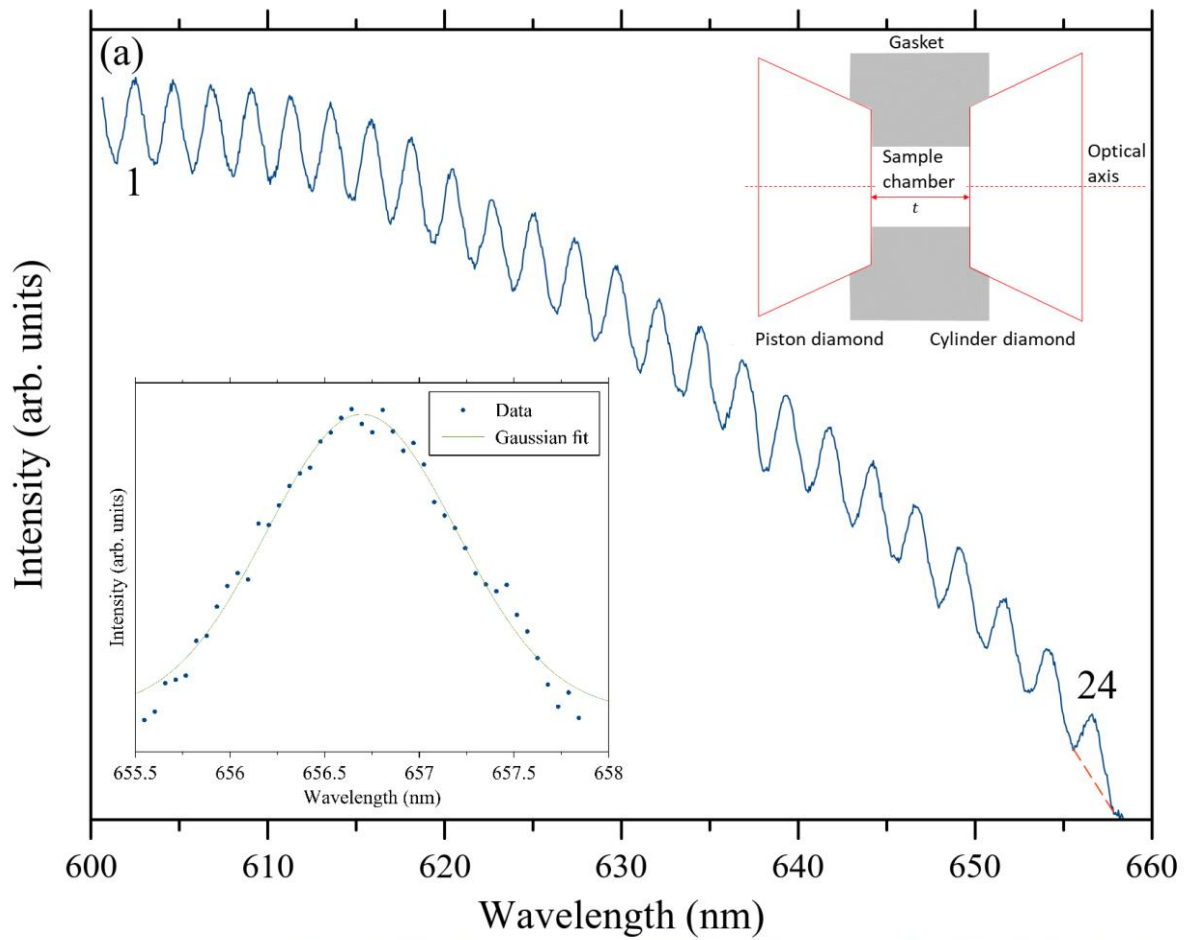

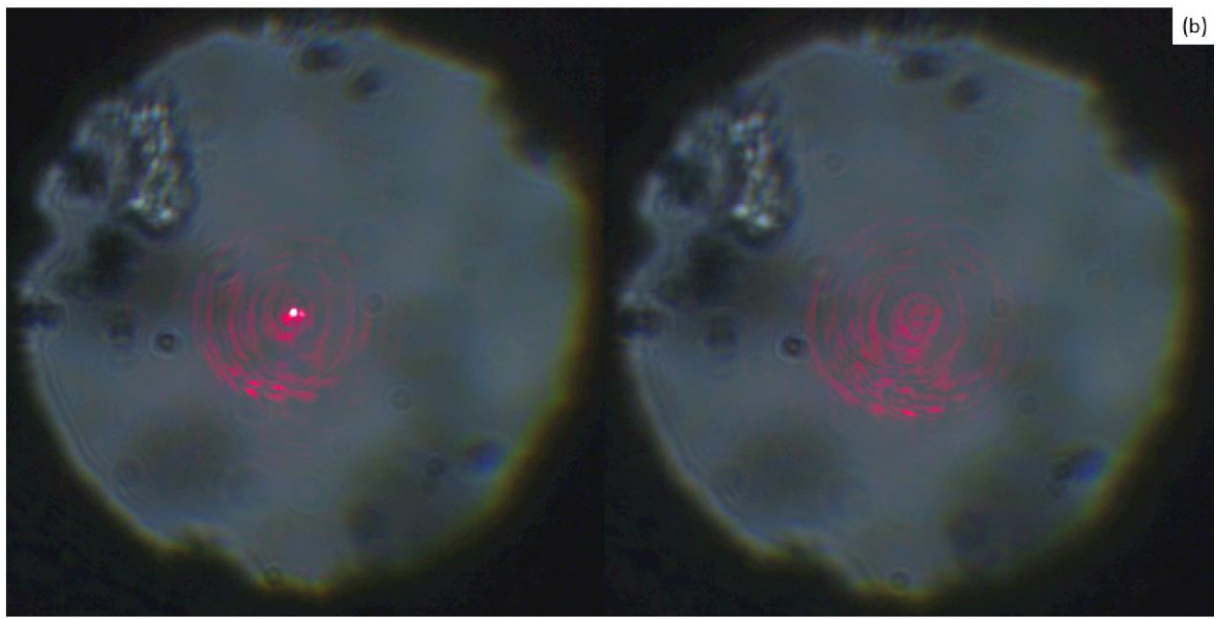

Figure 1. (a) Example white light fringes (insets showing Gaussian fit to the highest wavelength fringe after subtraction of the linear baseline shown and DAC geometry). (b) Images showing the laser focussed on one of the diamond culets (left) and defocussed part way between being focussed on the piston and cylinder diamond culets.

Evaluation of the errors arising in each measurement (shown in the supplementary material) reveals that the error in $t$ is dominated by the contribution arising from the ± 3 μm error in $t'$ (rather than the error in δ, typically ± 0.2 μm at high temperature and ± 0.05 μm at ambient temperature). Thus, if the RI were known independently, it would be possible to significantly reduce the error in $t$ by calculating it directly from $\delta = nt$ and discarding the measurement of $t'$. As noted earlier, there is a mutual dependence between the sample density $\rho$ and RI $n$. Experimentally, our measured value of the RI determines the measured $V_S$ and therefore the measured density. However, it is known from first principles (confirmed in a number of studies at ambient temperature and low temperature, reviewed later) that the sample density is by far the dominant factor in determining the RI, via the Lorentz-Lorenz law.

We propose (and test below) the following procedure to process our data, utilizing this mutual dependence. The procedure is applied to a set of data collected in a single experiment (i.e. there is no need for any normalization to account for having a different amount of sample in the chamber in successive experiments). To begin, we calculate $n$, $t$ and $V_S$ at all pressures in the dataset from δ, $t'$ and $A$ using equation (4). We then calculate the density in arbitrary units ($\rho = 1/V_S$). We plot $n$ vs $\rho$ and fit the Lorentz-Lorenz law [10] (formulated below as equation (5)) to the plot. In this fit there is a single adjustable parameter, the Lorentz-Lorenz factor $L$ (also in arbitrary units).

$$n = \sqrt{\frac{1 + 2L\rho}{1 - L\rho}}$$

(5)

We then use the RI values calculated using equation (5) with our fitted value of $L$ to recalculate $t$ using $\delta = nt$ (i.e. discarding the less accurate measurements of $t'$ in favour of the measurements of δ) and recalculate $V_S$. The process is then repeated until convergence in the values of $V_S$ obtained with each iteration. This takes 2 - 3 iterations for the datasets that we have collected.

3. Results
A. H$_2$O at 295 K

We conducted a control experiment on H$_2$O at 295 K, collecting 9 data points in the liquid state and one in the solid state, comparing our obtained RI values and $PV$ EOS to existing data in the literature. Figure 2 (a) shows our raw RI data as a function of our raw densities (in arbitrary units) and our fit using the Lorentz-Lorenz law (equation (5)). Aside from the outlying lowest density data point, the Lorentz-Lorenz fit is generally within the error on the experimental data points (including the highest density point in the solid state). Figure 2 (b) shows the raw RIs, and the RIs following two iterations of fitting with the Lorentz-Lorenz law, as a function of pressure. The RI of H$_2$O at ambient temperature has been the subject of a number of studies, reviewed in ref. [13]. Their power law fit, reproduced in our figure 2 (b), is a good representation of the experimental data in the literature.

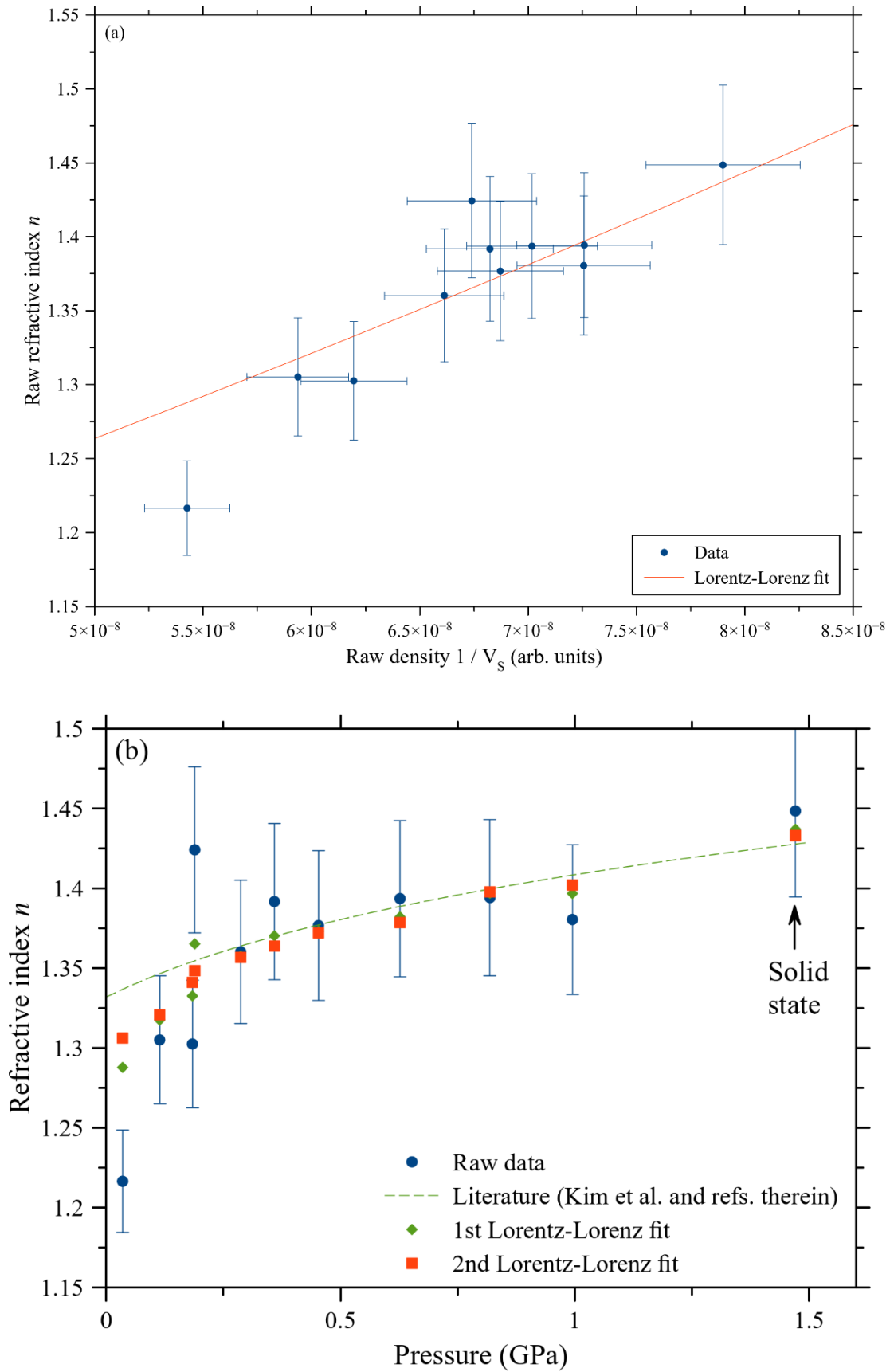

Figure 2. (a) Lorentz-Lorenz fit to the raw experimental data for $H_2O$ at 295 K. (b) RI values (raw, and after Lorentz-Lorenz fitting) plotted as a function of pressure. Error bars are calculated as outlined in the supplementary material. The error bars in pressure are too small to display (ca. 0.005 GPa).

In figure 3 (a) we show the raw volume data, and the volumes recalculated using the RI values following the second Lorentz-Lorenz fit. Our data are compared to the IAPWS fundamental equation of state, which is fitted to direct measurements of the PV EOS of $H_2O$ made in a piston-cylinder device [28]. Since the sample chamber volume is arbitrary, we have linearly rescaled our data to fit the IAPWS EOS. We also performed a third iteration of the Lorentz-Lorenz fit, however the change in the calculated volumes resulting from this was negligible. This is due to the fact that the scatter in the volume data following the second Lorentz-Lorenz fit closely matches the scatter in the sample chamber cross-section data (figure 3 inset). The cross-section data are used to calculate the densities for each Lorentz-Lorenz fit so any scatter in the cross-section data will affect the Lorentz-Lorenz fitting process as well as directly affecting the calculated volumes.

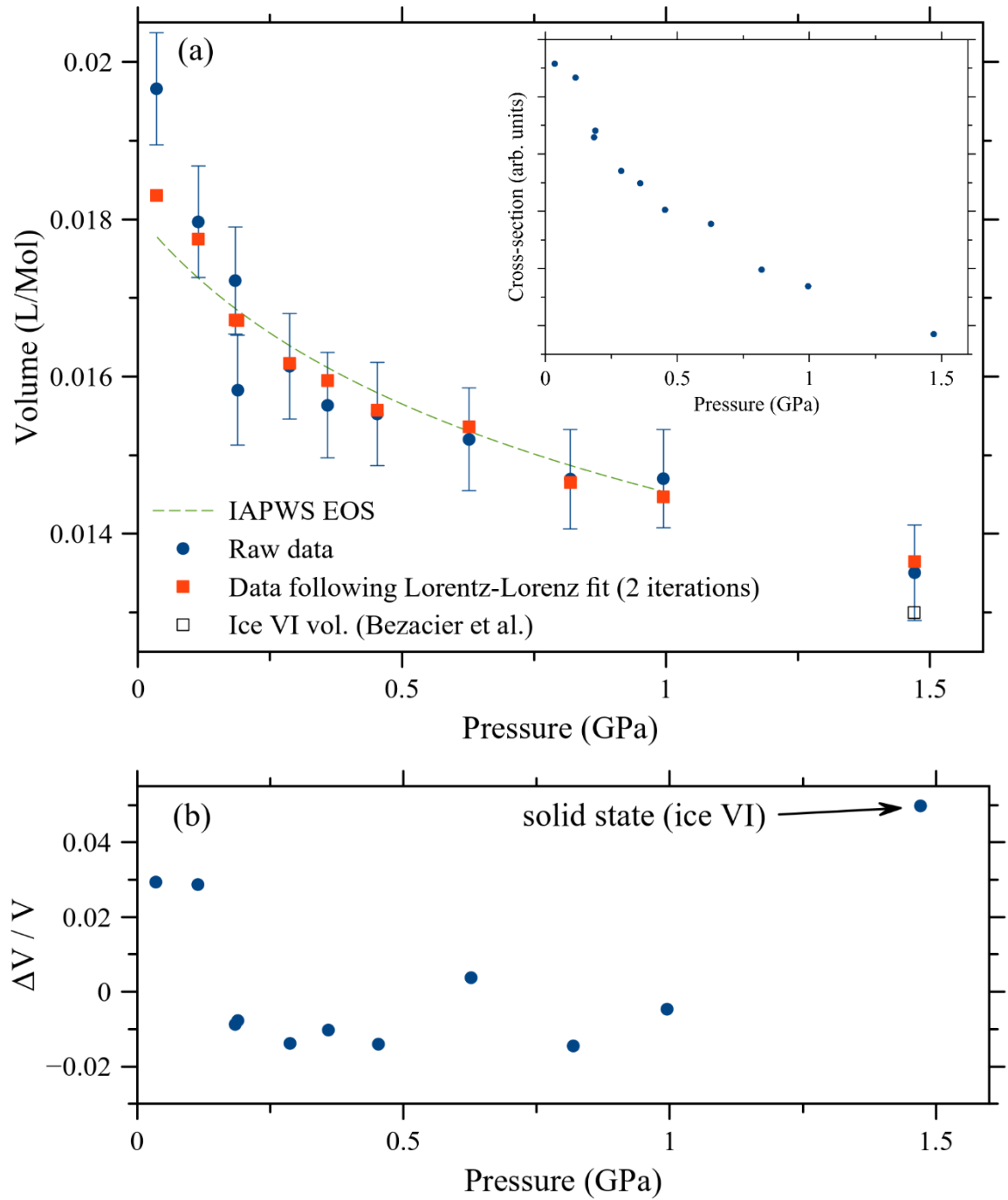

Figure 3 (a) Raw PV EOS data for $H_2O$, PV EOS data following two iterations of Lorentz-Lorenz fitting, output from the IAPWS EOS [28] and ice VI volume [29]. All measurements are at 295 K. (Inset) raw sample chamber cross-section data. (b) Difference plot comparing our volume data following two Lorentz-Lorenz fitting iterations to the IAPWS EOS and ice VI volume.

In figure 3 (b) we show the difference plot between our volumes (following 2 iterations of Lorentz-Lorenz fitting) and the literature (IAPWS EOS [28] in the liquid state and the ice VI volume [29] in the solid state). In general our data lie within ± 2% of the known EOS.

### B. KCl at 295 K

Our second control experiment was on KCl at 295 K. KCl exhibits a phase transition from the B1 phase to the B2 phase at ca. 2 GPa. Upon formation the B2 phase is opaque. However it becomes transparent upon further compression, allowing PV EOS measurement using optical methods. The PV EOS of the B1 and B2 phases has been accurately measured using synchrotron X-ray diffraction (ref. [30] and refs. therein). However, in contrast to $H_2O$, there is to our knowledge only a single experimental study of the RI pressure dependence across the phase transition [15].

Our results for KCl are shown in figure 4. Similarly to $H_2O$, only two Lorentz-Lorenz fit iterations (covering all the data across B1 and B2 phases) were required and further fits produced only negligible changes in the calculated volumes. The highest pressure datapoint (6.9 GPa) is clearly an outlier and the experiment was terminated at this point. This appeared to correlate with the emergence of significant nonhydrostaticity in the applied pressure (ruby line broadening) and has been noted in the other study of the pressure dependence of the RI of KCl [15].

Since we have the largest amount of data for the B2 phase, we scaled our sample chamber volumes following the Lorentz-Lorenz fit to obtain real atomic volumes by linearly scaling all B2 volumes to match the known EOS [30], then linearly scaling our B1 volumes by the same factor. The scaling procedure thus does not place any constraint on our measured volume change in the B1 → B2 transition.

Agreement with the known PV EOS (figure 4 (a) and (c)) is extremely good. In contrast to $H_2O$, the scatter in the PV EOS data after the Lorentz-Lorenz fit does not correlate with the scatter in the cross-section data. As is clear from the difference plot in figure 4 (c) there is a small systematic discrepancy between our measured volumes and the known EOS in the B1 phase.

In order to check our RI data, we digitized the data from ref. [15]. In doing so, we found two discrepancies in the data from this source. Firstly, the obtained value of the RI at ambient conditions was 1.480, a slight underestimate compared to the value now known at 632 nm (1.484 [31]). Secondly, if the change in the RI between the B1 and B2 phases is analysed using the X-ray diffraction EOS for KCl that is now known [30], it can be seen that it corresponds to a significant upward shift in the Lorentz-Lorenz factor $L$. This is not physically realistic as – if it changes it all – the Lorentz-Lorenz factor should decrease upon pressure (density) increase as this inhibits the polarizability of individual atoms.

We therefore constructed our own simple model for the RI of KCl, utilizing equation (5), using the known RI and density at ambient conditions to calculate the Lorentz-Lorenz factor, and assuming this remains constant. We used 62.40 Å$^3$ / atom and $n = 1.484$ (the known value at our laser wavelength of 632 nm [31]). We then used the known XRD EOS [30] to calculate the density, and hence RI, as a function of pressure for the B1 and B2 phases. Our experimental RI data (following Lorentz-Lorenz fitting) agree better with our model than with ref. [15].

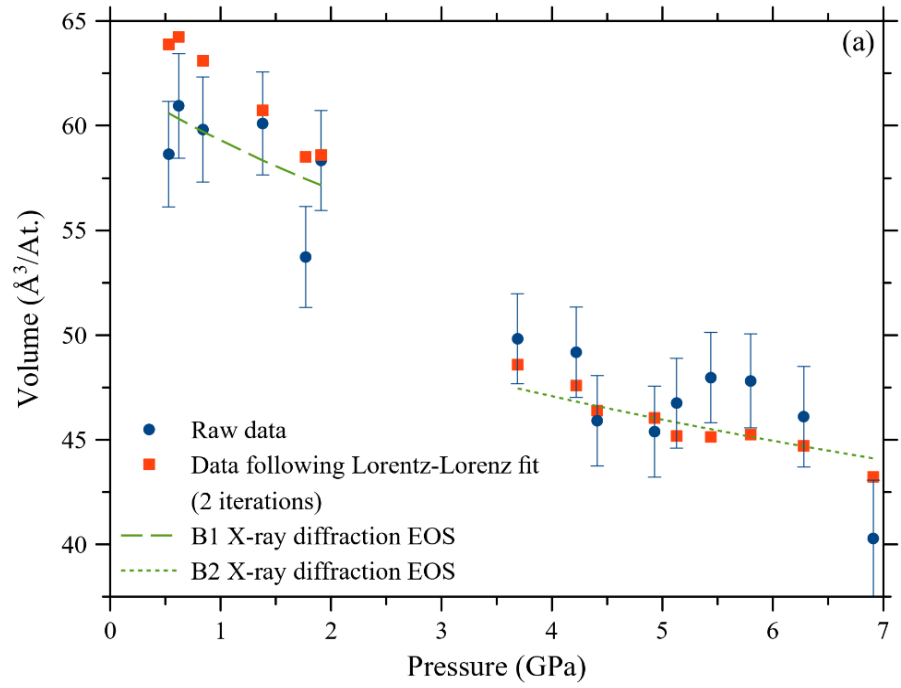

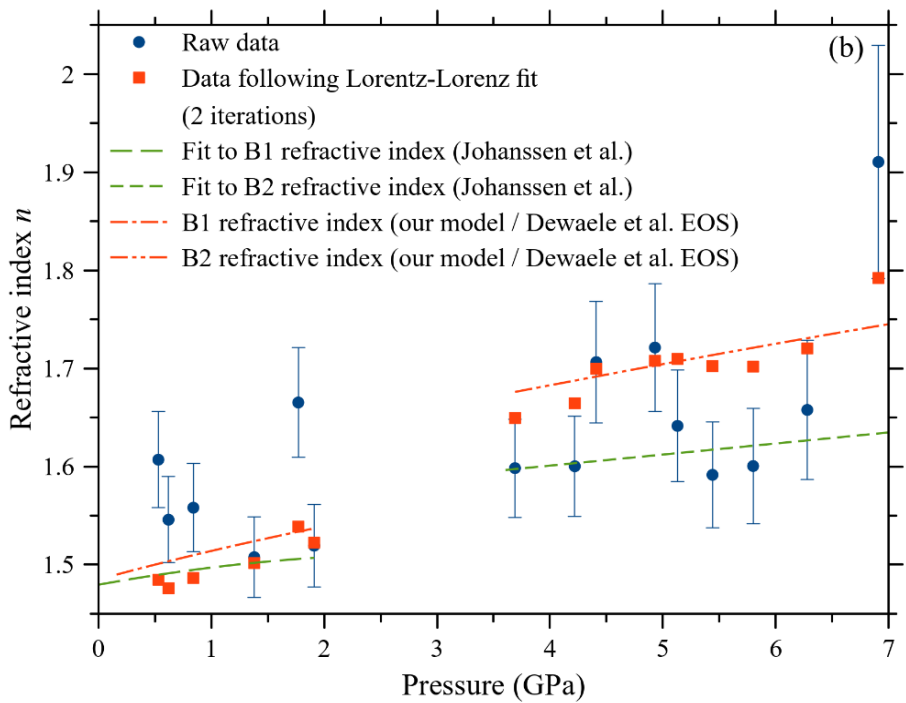

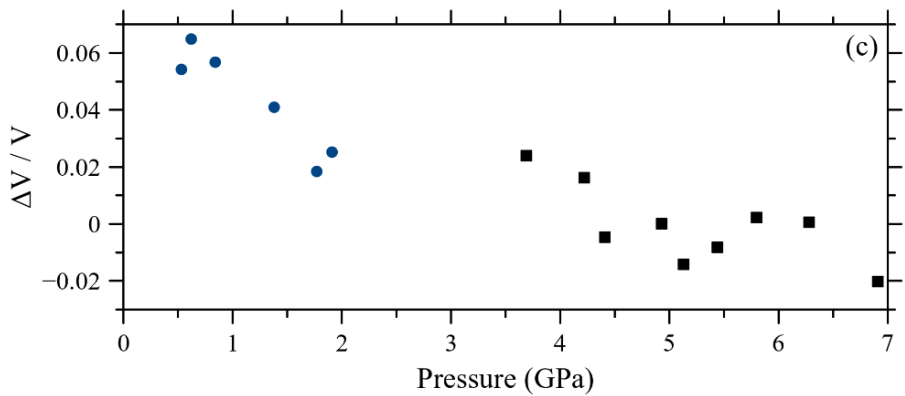

Figure 4. (a) Points: PV EOS for KCl at 295 K (raw data, and data following two rounds of Lorentz-Lorenz fitting). Lines: PV EOS known from X-ray diffraction data [30]. (b) Points: RI data for KCl at 295 K (raw data, and data following two iterations of Lorentz-Lorenz fitting). Lines: RI models as described in the text. (c) Difference plot comparing our PV EOS data (following two rounds of Lorentz-Lorenz fitting) to the PV EOS known from X-ray diffraction data [30].

### C. $CH_4$ at 514 K

We collected PV EOS data on liquid $CH_4$ at 514 K from 0.6 GPa up to the fusion curve, including a data point in the solid state. Here and subsequently, we will refer to our sample as a liquid despite it being above the critical temperature due to the fact that all experimental data collected are on the rigid liquid side of the Frenkel line (see refs. [5][32][33] and refs. therein). Similarly to the data collected in our control experiments described above, two iterations of fitting the RI data with the Lorentz-Lorenz law were required to achieve convergence. Our experimental data are shown in figure 5. At this temperature the data available in the literature are very limited. Our data agree with an extrapolation of the Wagner-Setzmann fundamental EOS for $CH_4$ [1] produced by the ThermoC software [34] (in the region where extrapolation is possible). The only other EOS measurement in existence in this PVT range, to our knowledge, is that resulting from the Brillouin spectroscopy study of Li et al. [8] including data at 539 K. In this study, the PV EOS was not produced using the iterative integration procedure used in other Brillouin studies and described in the introduction (equation (1)). Instead, the density was calculated directly from the RI using the Lorentz-Lorenz relation (equation (2)) with a value of the Lorentz-Lorenz factor $L$ obtained from earlier measurements of RI versus density at lower pressures and/or temperatures. There is thus an element of extrapolation in these data.

Furthermore, as is evident from figure 5, Li et al. observe no density change commensurate with freezing up to the highest pressure presented in their study, 5.12 GPa at 539 K. They state clearly that $CH_4$ did not undergo fusion to the solid state during any of their experiments. Within error, our assessment that the fusion curve was crossed between our 4.04(2) GPa and 4.94(2) GPa datapoints is consistent with the known fusion curve [35]. Li et al.'s assessment that $CH_4$ remains in the liquid state at 5.12 GPa, 539 K is not. There is a similar discrepancy between their data at other temperatures, including 298 K, and the known fusion curve. We have observed fusion ourselves at 300 K [33], at a pressure in agreement with ref. [35].

As is shown in figure 5, our PV EOS data are in reasonable agreement with those of Li et al. However, due to the discrepancy between Li et al.'s data and the known fusion curve, and the element of extrapolation in their density measurement, this agreement cannot be taken as a validation (or otherwise) of our data.

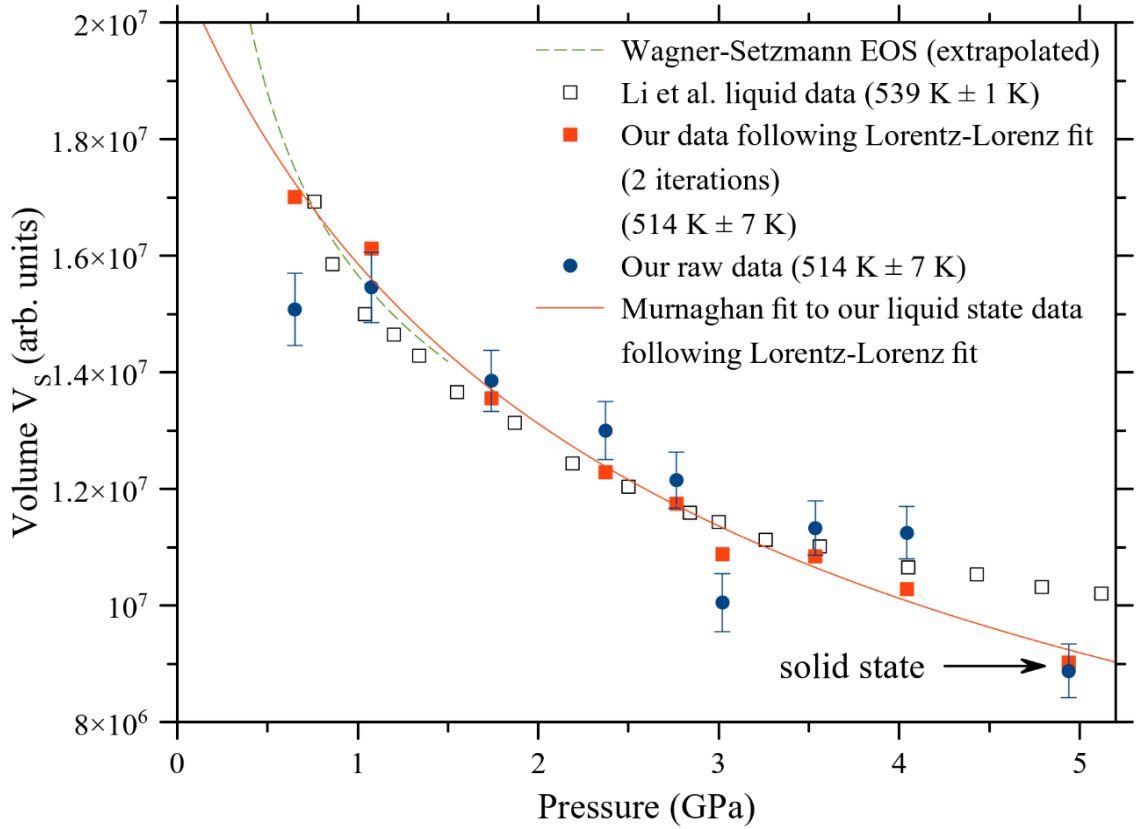

Figure 5. PV EOS of $CH_4$ at 514 K obtained from our data as described in the text, extrapolation of the Wagner-Setzmann EOS [1][34], and liquid volumes calculated from the RI obtained by Brillouin spectroscopy [8].

We completed our experiment on $CH_4$ at 514 K with a measurement of the volume in the solid state. In principle, this could be measured using X-ray or neutron diffraction. However, to our knowledge there are no EOS data available in the literature on solid $CH_4$ above ambient temperature. We understand that such measurements have been attempted, but have proved impossible due to the difficulty in forming a single crystal or a good powder from $CH_4$. Solid $CH_4$ tends to form a small number of crystals in the DAC, providing neither a single crystal nor a good powder comprised of many crystals [36][37].

The best verification that we can perform on our data is therefore to check that the volume change upon fusion displays (within error) an appropriate trend as a function of temperature. In ref. [5] we determined the volume change upon fusion at 300 K by extrapolating the known EOS in the liquid [1] and solid [38] states to the fusion curve. Both these EOS are backed by experimental data to pressures close to fusion. The fractional volume change $(V_{liq} - V_{sol})/V_{liq}$ was determined to be 0.066. To estimate the equivalent change at 514 K we performed a Murnaghan equation fit to our liquid state data and calculated the volume given by this Murnaghan fit at 4.94 GPa, the pressure of our solid state volume measurement. This gives a fractional volume change of 0.036 on fusion. This decrease in fractional volume change on increasing pressure seems reasonable. A more detailed analysis is probably not warranted due to the fact that we have only one datapoint in the solid state.

### D. H$_2$O at 509 K

We conducted two experiments on water at 509 K ± 2 K, covering the range from 0.09 GPa up to the fusion curve and comprising a total of 32 volume measurements in the liquid state. In this case, the Lorentz-Lorenz fitting procedure took three iterations to converge for each dataset. The PV EOS following this fitting are shown in figure 6.

The sample appears, at first sight, to exhibit anomalous behaviour of a small increase in compressibility in the pressure range between 1 – 3 GPa. This anomalous behaviour is likely to be the result of difficulties in pressure measurement during these experiments. Sm:SrB$_4$O$_7$ is the only pressure marker that is sufficiently accurate for this kind of work at 500 K (we know this from experience), however it has been known to react with hot water [26]. In our case this prevented any data collection above 1 GPa with Sm:SrB$_4$O$_7$ located inside the sample chamber. In both experiments we instead resorted to measuring pressure using a piece of Sm:SrB$_4$O$_7$ wedged between the gasket and diamond close to the pressure chamber above 1 GPa. The anomalously large jump in pressure at this point, that was not accompanied by a significant volume change (see figure 6 (c)), suggests that this led to an overestimate of the pressure. On the other hand, at 509 K H$_2$O freezes at 4.84 GPa according to the literature (ref. [5] and refs. therein). Comparison of this to the measured pressures of our highest pressure liquid state and lowest pressure solid state datapoint would seem to limit the error in our measured pressures to ± 0.3 GPa. However, any systematic error in pressure measurement would be caused by the water being soft compared to the gasket. This effect would be more significant at lower pressure. This, combined with the size of the anomalous pressure jump when we switched to the Sm:SrB$_4$O$_7$ outside the pressure chamber at ca. 1 GPa, would seem to indicate that ± 0.5 GPa is a more reasonable estimate of the potential error (random and/or systematic) in our pressure measurements above 1 GPa.

As far as the literature is concerned, it is unclear whether H$_2$O is expected to display any anomalous behaviour in this region. On the one hand, the pressures involved are far too high to observe the LDW → HDW transition (this has been observed in the vicinity of the water – ice I – ice III triple point at 209 MPa and should trend to lower pressure on temperature increase due to the entropy increase on the LDW → HDW transition). On the other hand, Kawamoto et al. observed a discontinuous change in the trend of Raman peak position vs pressure at around 1.5 GPa, at 573 K [39]. The PVT EOS produced by Abramson et al. using Brillouin spectroscopy exhibits no anomalous behaviour in this region [40], however the raw sound speed data collected up to crystallization at 473 K from which the EOS was derived are hard to fit with a single monotonic curve, and the data at 573 K are very limited. The same is true for the PVT EOS produced by Sanchez-Valle et al. [41] using Brillouin spectroscopy. In their speed of sound data at 573 K there appears to be a clear discontinuity in the 1 – 2 GPa region, however there is no discontinuity in the resulting PVT EOS.

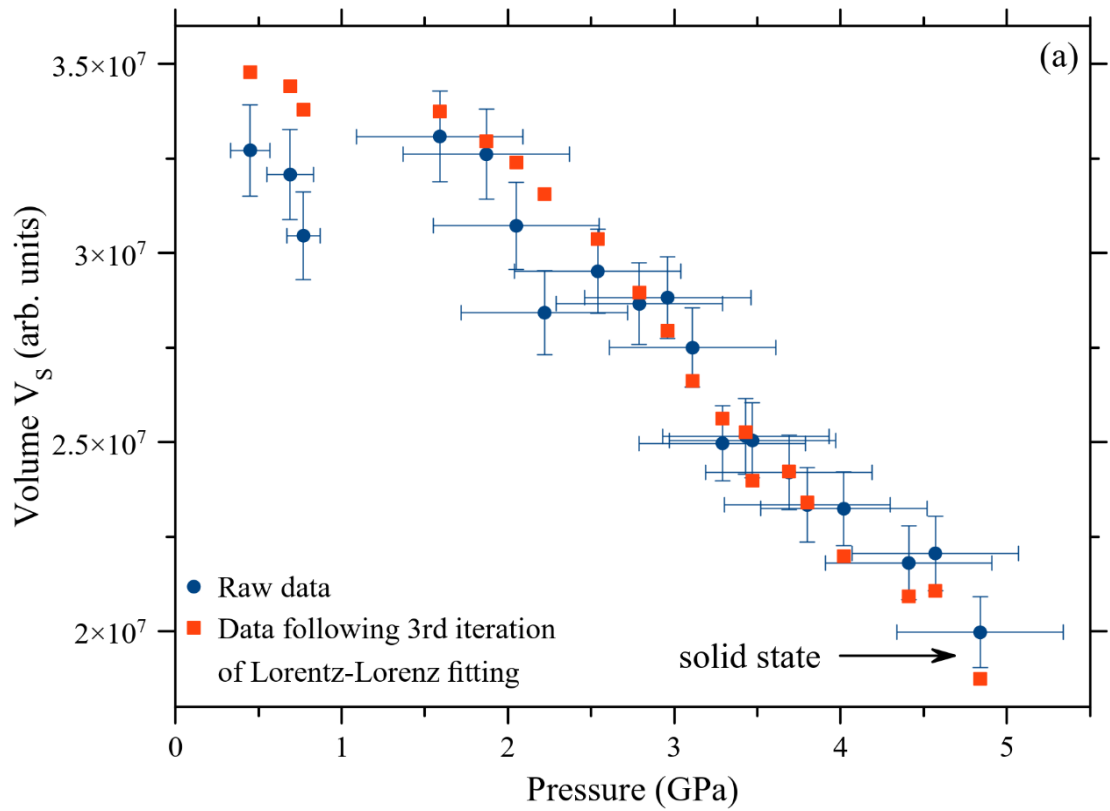

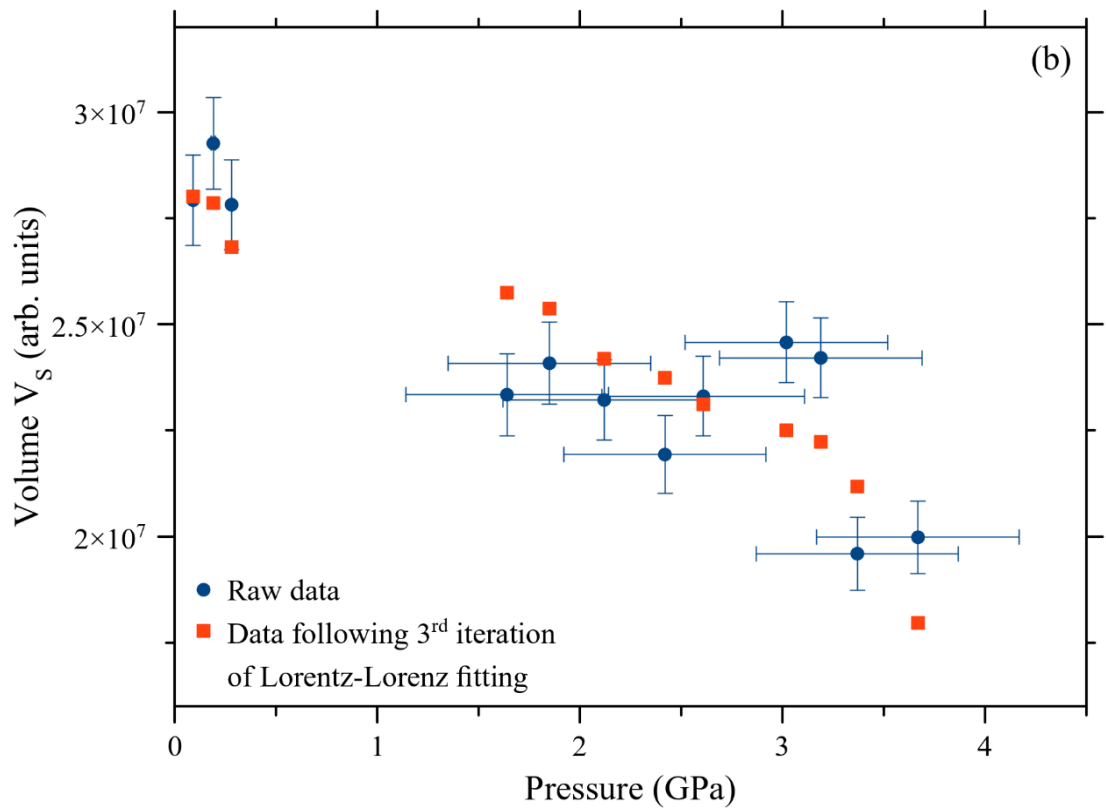

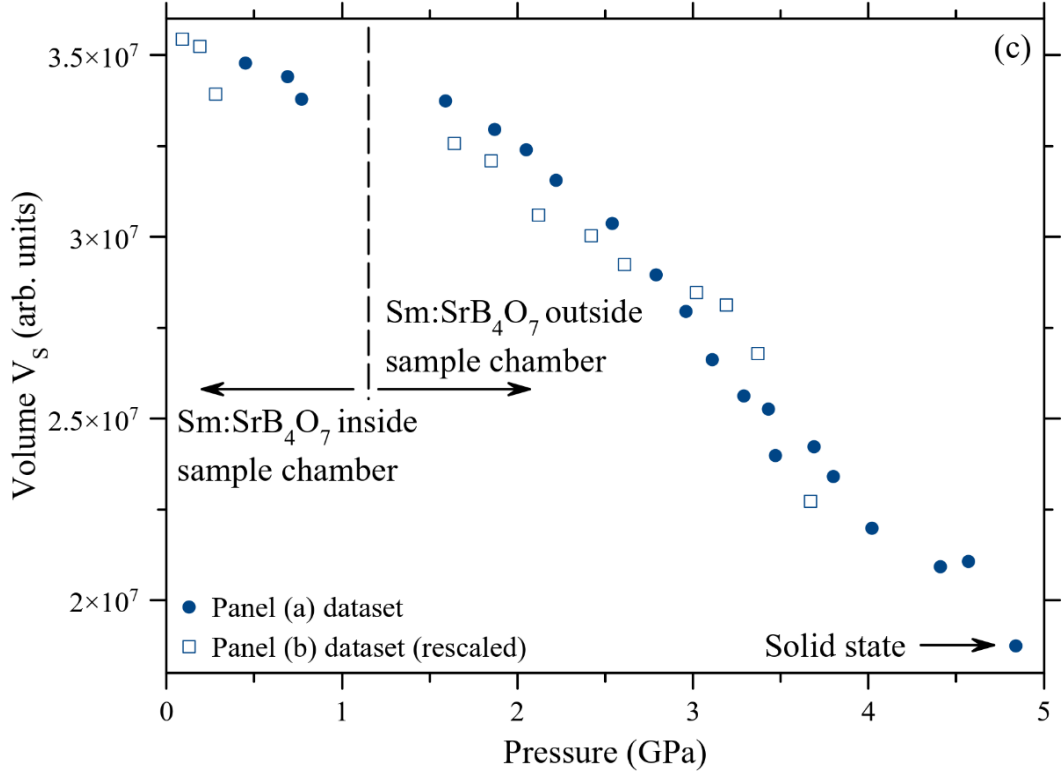

Figure 6. (a) and (b) Two separate datasets of raw volume data, and data following Lorentz-Lorenz fitting, plotted versus pressure for water at 509 K. (c) Volume versus pressure for both datasets following Lorentz-Lorenz fitting.

We are therefore unable, at this stage, to propose an EOS for water at 509 K. However, these datasets are extremely useful to test our iterative fitting procedure using the Lorentz-Lorenz law and (in section 3.E) a modified version thereof. Since this procedure utilizes the volume and RI measurements only, it is unaffected by errors in the pressure measurement.

Since we have measured the volume in the solid state (figure 6 panels (a) and (c), ice VII at 4.84 GPa, 509 K), a check on the effectiveness of our volume measurements can be made by comparing the volume change we have measured across the full pressure range in our work to the known change in volume between the liquid at 0.4 GPa and ice VII at 4.84 GPa. According to ref. [29] (Birch-Murnaghan EOS with temperature correction fitted to synchroton X-ray DAC EOS data), ice VII has a volume of 10.78 $cm^3$/Mol. at these conditions. Using this to convert our volume data from figure 6 (a) into units of $cm^3$/Mol., we can then check whether our EOS data agree with the known EOS at 509 K in the low pressure limit. The IAPWS EOS is backed by experimental data up to 0.4 GPa at this temperature, and gives a volume of 17.9 $cm^3$/Mol. at this point. At 0.45 GPa (our lowest pressure datapoint in this set) our volume measurement is 20.02 $cm^3$/Mol. This is as good an agreement as can reasonably be expected given that it hinges on a unit conversion made with a single datapoint.

### E. Experimental and fitting errors

Our experimental method consists, at each pressure, of three measurements: The sample chamber cross section $A$, the measurement from the white light interference pattern $\delta (= nt)$, and the distance moved by the DAC ($t'$) between the points where an incident laser beam entering the cell along the optical axis is focussed on the piston and cylinder diamond culets. A reasonable estimate of the experimental errors on these three measurements indicates that the error on the volume measurement is dominated by the large random error in $t'$. Our procedure of iteratively fitting the RI data with the Lorentz-Lorenz law eliminates this random error since the only use made of the $t'$ measurements is to provide the initial values for the RI for the first iteration of fitting.

However, the fitting procedure potentially introduces a systematic error since we assume the RI follows the Lorentz-Lorenz law. The Lorentz-Lorenz law is firmly rooted in well-understood theory of electromagnetism and is often used to calculate the fluid density from the RI [8][9][42][43]. However, the derivation of the Lorentz-Lorenz law does involve some assumptions [10]. Notably, it is assumed that the polarizability of the individual atoms or molecules comprising the sample remains constant. We would like to estimate the error arising from our choice to fit our RI data with this law. Fortunately, there are a lot of data available on samples in the liquid and solid states at ambient temperature, and at low temperature, that can enable a reasonable assessment to be made of how well this law is obeyed.

We have undertaken such an exercise for $CH_4$. In figure 7 the data are shown for RI versus density extracted from a range of studies in the liquid, solid and gas states covering temperatures from 14 K to 373 K and pressures from 0.1 MPa to 12 GPa [8][42][44][45][46][47][48]. Details on data preparation are given in the supplementary material. At the very highest densities studied (covered only by Hebert et al.'s study of the solid phase A [45]) there appears to be, at first sight, a small systematic discrepancy between the measured RIs and the Lorentz-Lorenz law fit, with the measured RIs trending to slightly smaller values at high density than would be expected according to the Lorentz-Lorenz law. This is what could be expected at high density since this will inhibit the polarizability of individual atoms, compromising the assumption made in the derivation of the Lorentz-Lorenz law.

However, there is considerable scope for systematic errors to arise between the 10 datasets that are shown. A variety of different methods are used in these data to measure RI, pressure, temperature and density. From examination of the data throughout figure 7 it is clear that there is a systematic discrepancy between the Hebert et al. liquid and phase I datasets [45] and other datasets at similar densities, that can be only attributed to a systematic error. When these two datasets are excluded from the Lorentz-Lorenz law fitting procedure, the systematic discrepancy at high density that we tentatively attributed to the failure of the Lorentz-Lorenz law in the previous paragraph disappears. We thus conclude that there is at present no convincing experimental evidence for any failure of the Lorentz-Lorenz law, or variation in the value of the Lorentz-Lorenz factor $L$ with density, in $CH_4$.

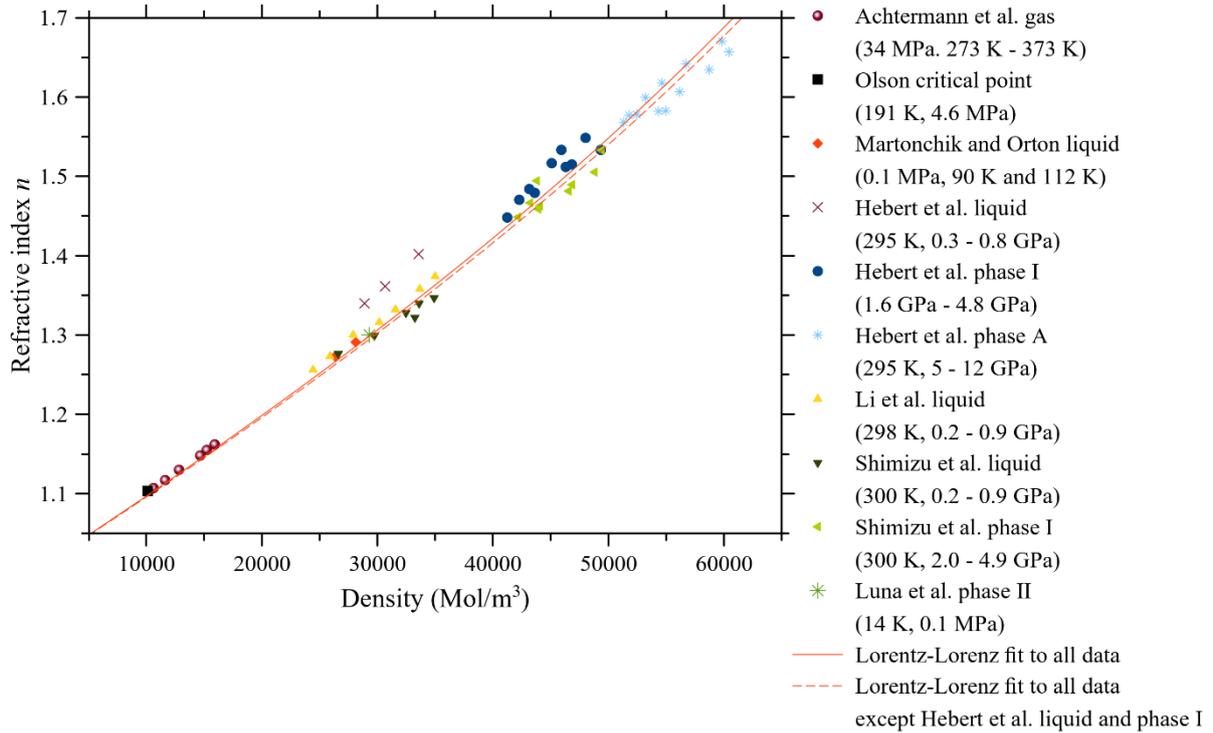

Figure 7. (points) CH$_4$ RI measurements from the literature [8][42][44][45][46][47][48] plotted as a function of density. (solid red line) Lorentz-Lorenz fit to all data. (dashed red line) Lorentz-Lorenz fit with Hebert et al. liquid and phase I data excluded.

In ref. [12], Dewaele et al. reviewed the available RI data for fluid and solid H$_2$O up to ambient temperature using similar methodology to our review for CH$_4$, and concluded that the assumption of a constant Lorentz-Lorenz factor would lead to an underestimate of the RI of about 3% at 35 GPa at ambient temperature, a significantly higher pressure and density than that reached in our study. They showed that the Lorentz-Lorenz factor decreases linearly with density, in which case the RI data should be fitted using the following equation in place of equation (5), with the constraint that $L' < 0$:

$$n = \sqrt{\frac{1 + 2(L_0 + L'\rho)\rho}{1 - 2(L_0 + L'\rho)\rho}}$$

(6)

We attempted the Lorentz-Lorenz fitting with this modification for all three of our H$_2$O datasets. For the 295 K dataset, and for the first of the 509 K datasets, the fitting process converged rapidly to $L' = 0$. Only for the second dataset at 509 K did the fit using equation (5) converge to a non-zero value for $L'$. In this case, 17 of the 20 RIs calculated using equation (6) lay within +1.4% and -0.8% of the RIs calculated using equation (5). The outliers were the 3 highest pressure datapoints. Considering this dataset along with the other two datasets in which there was no measurable discrepancy between the RIs and volumes obtained using equations (5) and (6) we propose ± 1% as a reasonable estimate of the error in the RI introduced by the choice of fitting process.

To conclude this section, we will examine how the potential error resulting from the choice of equation (5) to fit the RI data combines with the experimental errors to give the overall error in our calculated volumes. From equation (4) we obtain $V_S = A\delta/n$. Thus the fractional error in $V_S$ is obtained simply by adding the fractional errors in $A$, $\delta$ and $n$ in quadrature. We have estimated the fractional error in $A$ to be ± 2.5% (see supplementary material), and the fractional error in $\delta$ is negligible in comparison.

Even in the high temperature data where the fringes become significantly fainter than those shown in figure 1, counting over a significant number or fringes plotted as a function of wavenumber keeps the fractional error in $\delta$ down to ± 0.2% typically. The determination of the RI using the Lorentz-Lorenz law fitting (with a ± 1% error as proposed above) replaces the direct measurement of $t'$, which has an estimated fractional error varying between ± 2% and ± 5% (see earlier sections and supplementary material). Thus – according to our best efforts to quantify the known errors in our measurements and fits – the error in the volumes we have determined using the Lorentz-Lorenz fitting process is at present about ± 2.7%. The errors measured in our control experiments are consistent with this estimate: The standard deviation of the difference plot data from figures 3 (b) and 4 (c) is 2.5%. Accounting for the errors in $t'$, $\delta$ and $A$ that contribute to the errors in the raw values of $V_S$, and the ± 1% error in $n$ in the fitted values of $V_S$, we can see that the error margins in the raw $V_S$ values and $V_S$ values following Lorentz-Lorenz fitting generally overlap.

## 4. Conclusions

We have presented here a new method for determining the PV equation of state of fluids and transparent solids in the DAC. Our method is novel in two ways: Firstly, the combination of white light interference with confocal microscopy to measure an EOS. Secondly, our analysis method that exploits the need for the observed trends in density and RI to be mutually consistent.

A reasonable estimate of the margins of error in our experimental measurements and analysis method indicates that the error in our calculated volumes is about ± 2.7%. Furthermore, it indicates how the error in the calculated volumes can be reduced in future work. The error in the white light interference measurement leading to the value of $\delta$ rarely exceeds ± 0.2%. It is unlikely that this can be reduced further, or that it will become the dominant source of error. The error in the cross-section measurement is currently the dominant error at ± 2.5%, and there is scope to significantly reduce this in future experiments with an improved microscope. In ref. [14], Lobanov et al. measured DAC sample chamber cross-sections at ambient temperature and found that the error in the cross-section measurement gradually increased with pressure, reaching ± 1.5% at 111 GPa. On this basis, it would seem feasible to reduce the cross-section measurement error in the pressure range relevant to this work (< 10 GPa) to significantly below ± 1%.

The error introduced in the process of fitting $n(\rho)$ to equation (5) or (6) could also be reduced significantly. In particular, for most fluids and transparent solids that are of interest, accurate RI data already exist at lower pressures (for instance measurements in the dense gas state at ambient pressure, in the critical region and in the liquid and solid states at low temperature, ambient pressure). In the present work we have sought to explore what can be achieved without utilizing any *a priori* knowledge of the RI (except for assuming $n = 1$ in vacuum). However, incorporating *a priori* knowledge of the RI (especially where the experimental errors are small compared to the errors in our raw values of the RI) could assist significantly.

It is evident from inspection of our figures that our data contain occasional clear outliers. It would be desirable, in future, to increase the density of datapoints so that outliers can be identified and removed in an objective manner based on statistical tests.

Finally, there is scope to reduce the ± 3 μm error in $t'$ which would ultimately lead to tighter constraints on the outcome of the fitting process. At present, the focal points leading to the measurement of $t'$ are located by visual observation, and movement of the stage is via manual micrometer. So, whilst position can be measured to a precision of 1 μm, it cannot necessarily be controlled to this precision.

In a future apparatus design a professionally constructed confocal microscopy setup with automatic stage or lens movement (perhaps similar to that described in work at ambient temperature in refs. [17] and [18]) could reduce this error significantly.

The measurements presented here were conducted on a modified photoluminescence / Raman spectrometer. It seems feasible, from the arguments presented above, that conducting future measurements on apparatus purpose-designed for these experiments could allow the error in the volume measurements to be reduced to ca. ± 1% using the methodologies outlined in the previous two paragraphs. This is similar to the error in synchrotron X-ray diffraction measurements of volume in the solid state.

On a conceptual level also, our work brings high temperature fluid EOS measurement in the DAC far closer to what is possible for solids using synchrotron X-ray diffraction. To produce Brillouin scattering EOS data (the current state-of-the-art for fluids in the DAC) there are usually several empirical fitting stages as it is hard to do the numerical integration with the adiabatic → isothermal correction without these. For instance, the RI data and/or speed of sound data can be fitted with an empirical function of temperature and/or pressure. Our method only includes one intermediate fitting stage, that when the RI data are fitted with the Lorentz-Lorenz law. This fitting is not empirical. This brings us significantly closer to the state-of-the-art for solids, in which the final processing stage of fitting the $V(P)$ graph with an analytical EOS is the only semi-empirical part of the process.

Similarly, in Brillouin scattering it is necessary to numerically integrate as a function of pressure/volume/density to actually produce an EOS. So an error at one pressure affects all subsequent (higher) pressures. Our method, similarly to synchrotron X-ray diffraction on solids, contains no such weakness. In addition, our method fits to the density (rather than pressure) so can work across phase transitions.

In earlier work at ambient temperature utilizing the Fabry-Perot fringes method [12], it has been proposed that the hard limitation on the pressure at which EOS can be measured using optical methods such as ours is the phenomenon of "cupping" of the diamonds, i.e. the diamond culet bending outwards into the sample chamber. This is also our view, based on our experience gained in the present work. The "cupping" effect causes two errors. Firstly, the reduction in the degree of parallelism between the diamond surfaces reduces the signal-to-noise ratio of the fringes. The error resulting from this phenomenon can be quantified (and we have done so): As long as one can measure across a reasonable number of fringes, the error in δ resulting from the poor signal-to-noise ratio usually remains small at about ± 0.2%.

Secondly, the "cupping" effect ensures that any calculation of sample chamber volume based on $V = At$ is subject to an error which is very hard to quantify if $t$ is varying across the sample chamber area. In future, if it is possible to reduce the error in $t'$ to ca. ± 1 µm, it will be possible to quantify the error due to the "cupping" effect by repeating the $t'$ measurement at different locations in the sample chamber – or least put an upper limit on it.

It is worth noting that the "cupping" effect also leads to a hard limitation on the pressures at which EOS can be measured using Brillouin spectroscopy. All measurements of sound speed in the DAC require knowledge of the scattering geometry – based on the assumption that the diamond culets remain parallel to each other and perpendicular to the optical axis. This is no longer the case if "cupping" is taking place. For geometric reasons, the symmetric forward scattering geometry in Brillouin spectroscopy is probably the most vulnerable to errors introduced by "cupping" and the 180° back-scattering geometry probably the least vulnerable. This is unfortunate, as the symmetric forward scattering (platelet) geometry is the only geometry in which the sound speed can be calculated without knowledge of the RI [22]. Clearly, in future work the combination of our new methodology to calculate the RI and EOS with Brillouin spectroscopy in the 180° backscattering geometry (utilizing the RI calculated with our methodology) to produce an independent EOS determination could be extremely powerful, producing well-constrained values for the heat capacity as a by-product using equation (1).

Finally, our work has application in the study of the fusion curve and high temperature solid EOS. The volumes on each side of the fusion curve can be determined using the same method, and EOS can be obtained for high temperature solids for which EOS measurement using synchrotron X-ray diffraction cannot be performed due to the impossibility of obtaining a good powder or single crystal. However, it is clear – looking at all our datasets above – that there is some scatter in the data produced using our methodology. It would therefore be essential to collect a significant number of datapoints in the solid state, as well as in the liquid state, to account for this.

**Supplementary material**

The supplementary material contains the derivation of equation (4), full error calculations, examples of white light fringe data at high temperature, plots of refractive index versus pressure for $C_2H_6$ at 380 K and $C_3H_8$ at 295 K, EOS data for NaCl at 295 K, further experimental details, and details of how the data in figure 7 were prepared.

**Acknowledgements**

We would like to acknowledge Alex Ritchie, Ciprian Pruteanu and Alan Soper for useful discussions, Simon Hunt for useful discussions and critical feedback on a draft of the manuscript, Frédéric Datchi for provision of a Strontium Borate sample, and Mike Clegg for construction of diamond anvil cells in the University of Salford workshop.

**Author contributions**

JEP designed the study, performed experiments, analysed data and wrote the paper. CEAR, LJJ, JP, KW and YD performed experiments and analysed data, BM analysed data.

**Data availability**

The data that support the findings of this study are available from the corresponding author upon reasonable request.

**Conflicts of interest**

The authors have no conflicts of interest to disclose.

# Supplementary information

**Contents**

**Derivation of equations to calculate $n$ and $t$ from $\delta$ and $t'$**

Here we derive equation (4) in the main text. Figure S1 outlines the geometry and nomenclature used.

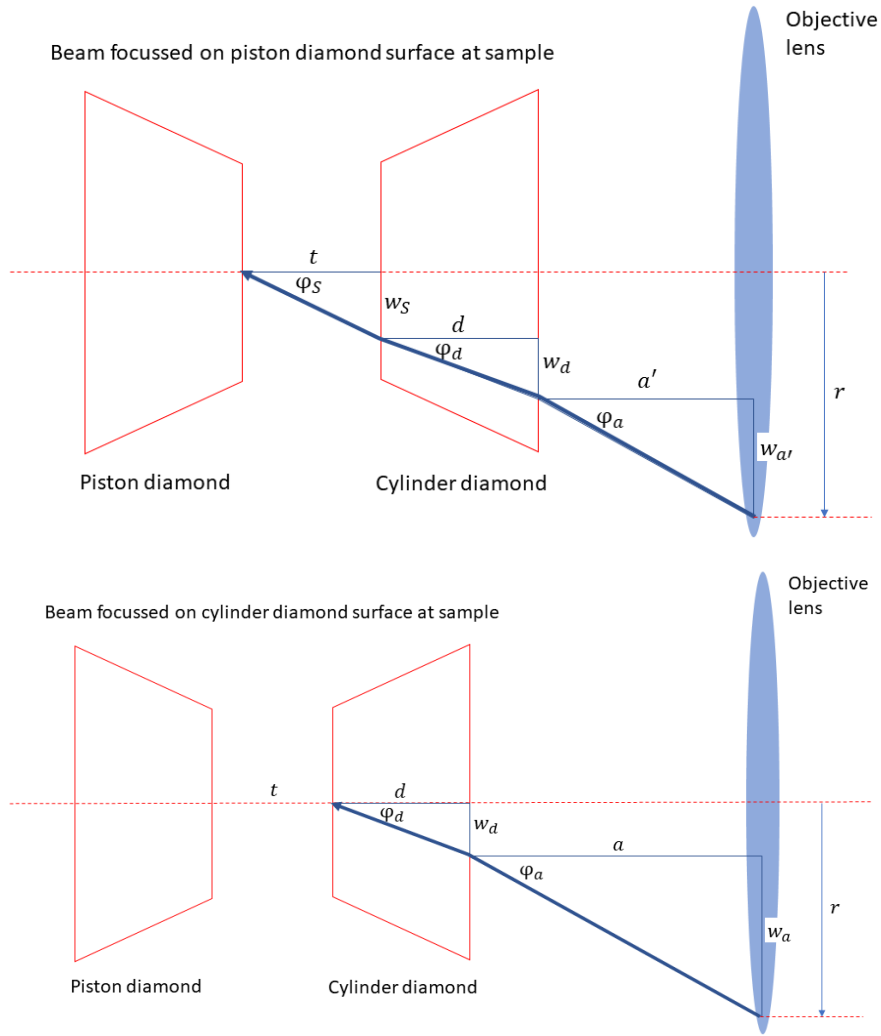

Figure S1. Geometry and nomenclature for focussing of collimated laser beam on the culets of the cylinder (top) and piston (bottom) diamonds.

The incident collimated beam is of radius $r$, and a ray (shown as thick blue line) from the edge of the beam is incident towards the focal point at an angle $\varphi_a$ in both cases, before being refracted at the air – cylinder diamond interface and (in the case where the beam is focussed on the piston diamond) at the diamond – sample interface.

Applying Snell's law to both sides of the cylinder diamond in the case where the beam is focussed on culet of the piston diamond leads to (using the paraxial small angle approximation):

$$n\varphi_s = \varphi_a$$

(S1)

Equation (S2) is obtained by adding up the triangle sides equal to $r$ on each diagram:

$$w_a = w_s + w_{a'}$$

Equations (S3) and (S4) are obtained by applying the small angle approximation to the relevant triangles.

$$\varphi_s = \frac{w_s}{t}$$

(S3)

$$\varphi_a = \frac{w_{a'}}{a'}$$

(S4)

In equations (S1) – (S4) the values of the following variables are known (or at least, we could measure them in principle if we wanted to): $\varphi_a, w_a$. The values of the following six variables are unknown: $n, \varphi_s, w_s, t, w_{a'}, a'$. We would like to calculate $t$ and $n$. We will now introduce our two experimental measurements. The white light interference experiment gives us the product $nt$ (we label this as $\delta$):

$$\delta = nt$$

(S5)

The second measurement is the distance the DAC is moved between the location where the beam is focussed on the cylinder diamond – sample interface, and the location where it is focussed on the piston diamond – sample interface. This measurement (made with the micrometer attached to the sample stage) will be a length similar to $t$. We will label it $t'$. It is easiest to visualise this as the difference in distance from the (fixed position) lens to the back of the cylinder diamond between the 2 diagrams above. In this case we obtain:

$$t' = a - a'$$

(S6)

This equation has introduced a variable ($a$) not present in equations (S1) – (S5). However, $a$ could be measured just like $\varphi_a, w_a$. In any case we now have 6 equations (S1) – (S6) and 6 unknown variables ($n, \varphi_s, w_s, t, w_{a'}, a'$), so we can solve for $n$ and $t$:

$$t = \sqrt{\delta t'}$$

$$n = \sqrt{\frac{\delta}{t'}}$$

(S7)

During this process $\varphi_a, w_a, a$ all cancel out. If we relax the paraxial approximation (i.e. write $n \sin \varphi_s = \sin \varphi_a$ in place of equation S1 etc.) this is no longer the case. Physically, if $w_a$ and $\varphi_a$ did *not* cancel out, this would correspond to the focal point becoming smeared out along the optical axis (longitudinal aberration). This phenomenon was studied in ref. [27] and, in the next section, we use their methodology to quantify the potential for systematic error arising from longitudinal aberration.

## Longitudinal aberration caused by the air-diamond and diamond-sample interfaces

In ref. [27], Ghatak studied the longitudinal aberration caused by insertion of a plane refracting surface into a beam between an objective lens and the focal point. They showed that (even in the absence of any aberration caused by the objective lens) when the paraxial approximation is relaxed, the focal point shifts according to how far out from the optical axis the incoming ray strikes the refracting surface. Here we will write their principal result using our notation and apply it to calculating the magnitude of the systematic error caused by our adoption of the paraxial approximation.

Considering the case where a paraxial ray is focussed on the culet of the cylinder diamond (figure S1 lower panel), accounting for non-paraxial effects to first order would shift the focal point by an amount $\Delta f_1$ depending on the value of $w_d$ (where $w_d = d \tan \varphi_d$):

$$\Delta f_1 \approx \frac{[d \tan \varphi_d]^2}{2d \left(\frac{1}{n_d}\right)} \left[\left(\frac{1}{n_d}\right)^2 - 1\right]$$

(S8)

Here $n_d$ is the refractive index of diamond, the other symbols are as defined in figure S1 and $\Delta f_1$ is defined as being positive if the focal point shifts away from the objective lens. Applying Snell's law to the air-diamond interface ($n_d \sin \varphi_d = \sin \varphi_a$) allows us to calculate the longitudinal aberration in this focal length $\Delta f_1$ as a function of known quantities $n_d, d$ and $\varphi_a$ under the approximation $\Delta f_1 \ll d$. We will assume that the angles take the maximum values permitted by the numerical apertures (NAs) of the lens and DAC.

Now consider the case where a paraxial ray is focussed on the culet of the piston diamond (figure S1 upper panel). To first order, we can account for longitudinal aberration to this beam caused at the air-diamond interface ($\Delta f_2$) by making the adjustment to equation (S8) $d \rightarrow d + t$:

$$\Delta f_2 \approx \frac{[(d+t) \tan \varphi_d]^2}{2(d+t) \left(\frac{1}{n_d}\right)} \left[\left(\frac{1}{n_d}\right)^2 - 1\right]$$

(S9)

The longitudinal aberration to this beam caused by refraction at the diamond-sample interface ($\Delta f_3$) is given (using the same methodology and sign convention as equations (S8) and (S9), and the same notation as figure S1) by:

$$\Delta f_3 \approx \frac{[t \tan \varphi_s]^2}{2t \left(\frac{n_d}{n}\right)} \left[\left(\frac{n_d}{n}\right)^2 - 1\right]$$

(S10)

The angles $\varphi_s$ and $\varphi_d$ can, provided the refractive index $n$ of the sample is known, be calculated directly from $\varphi_a$ using Snell's law. The overall systematic error ($\Delta t'_{ab.}$) due to the combined effect of longitudinal aberrations when the laser is focussed on the piston and cylinder diamonds is therefore estimated as:

$$\Delta t'_{ab.} \approx \frac{\Delta f_3 + \Delta f_2 - \Delta f_1}{2}$$

(S11)

Here, the factor of 2 is present because we would expect the maximum in laser intensity to be part way between the paraxial focal point and the non-paraxial focal point where the angles are at their maximum values permitted by the numerical apertures of the lens and DAC. This is, of course, an estimate. An exact calculation would require measurement of the beam profile. In figure S2 (below), we plot $\Delta t'_{ab.}$ as a function of $\varphi_a$ for typical parameters $n = 1.5$, $t = 75$ µm, $d = 1.6$ mm, and the known refractive index of diamond ($n_d = 2.4$). At the opening angle utilized in this work (corresponding to a lens NA of 0.3) we estimate the % error in $t'$ resulting from use of the paraxial approximation to be 0.35%. This corresponds to an absolute error in $t'$ of 0.17 µm.

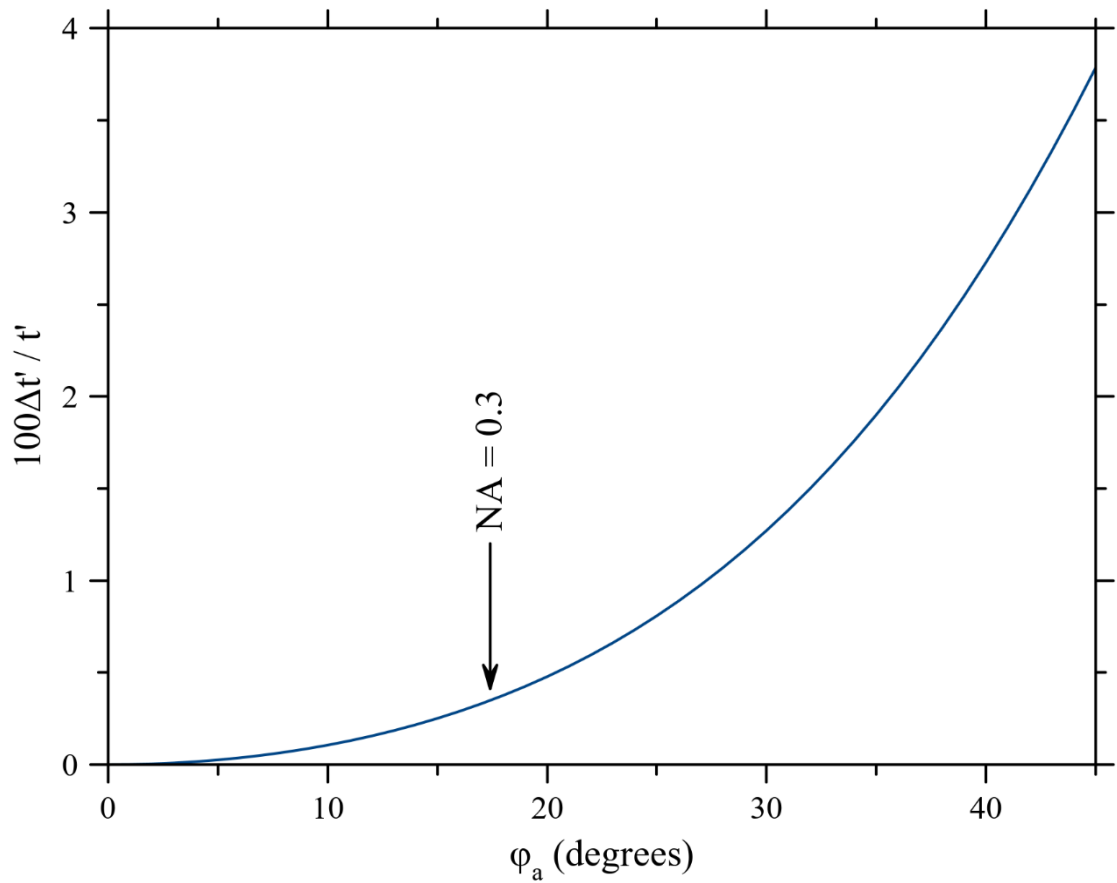

Figure S2. % error in $t'$ resulting from use of the paraxial approximation estimated using the methodology above, plotted as a function of angle $\varphi_a$. The present work utilized an objective lens with NA = 0.3, equivalent to $\varphi_a = 17.45°$ maximum.

**Derivation of equation (3) in the main text**

Inside the DAC sample chamber, light with a free-space wavelength of $\lambda$ has a wavelength of $\lambda/n$. When we shine white light through the DAC along the optical axis, the interference fringes therefore occur at wavelengths for which $\lambda l = 2nt$ where $l$ is an integer. The peaks are equally spaced as a function of wavenumber and for a set of $p$ peaks where the peaks at each end of the set have wavelengths $\lambda_{A,B}$ we can write:

$$\frac{1}{\lambda_A} - \frac{1}{\lambda_B} = \frac{1}{2nt}[l + (p-1) - l]$$

(S12)

From which equation (3) in the main text trivially follows with $\delta = nt$. Here we assume there is no dispersion in our samples. Experimentally, we also ensured for all samples except $H_2O$ at 509 K that the wavelength of the laser used for the determination of $t'$ (632 nm) lies within the range of fringes utilized. In the $H_2O$ samples at 509 K we could only obtain sufficiently good quality fringes to fit in the range 575 – 620 nm. However, we could detect no dispersion within this range. No significant dispersion would be expected in any case, as discussed in ref. [12].

**Error calculations**

The value of δ is obtained using equation (3) in the main text with $\lambda_{A,B}$ obtained from Gaussian fits to the relevant fringes after linear background subtraction. The errors in these wavelengths are given by the fitting software (Magicplot Pro) and (adding errors in quadrature in the normal manner) propagate through to give the following error in δ:

$$\Delta\delta = \frac{p-1}{(\lambda_B - \lambda_A)^2}\sqrt{\lambda_B^4(\Delta\lambda_A)^2 + \lambda_A^4(\Delta\lambda_B)^2}$$

(S13)

Inspection of the derivation for equation (3) in the main text shows that the fringes are equally spaced as a function of wavenumber rather than wavelength. In the high temperature data we found it easier to obtain δ from a plot of intensity versus wavenumber for this reason. In this case, the error in δ is calculated from errors in the relevant wavenumbers ($\nu_A = 1/\lambda_A$ etc.) using:

$$\Delta\delta = \frac{p-1}{(\nu_B - \nu_A)^2}\sqrt{(\Delta\nu_A)^2 + (\Delta\nu_A)^2}$$

(S14)

The error in δ is insignificant at ambient temperature, and at lower pressures at high temperature (sometimes as low as ± 0.001 µm). At higher pressure combined with high temperature it can become significant (ca. ± 1 µm occasionally), but not dominant. The increase as a function of pressure means that it is essential to calculate the error using the equations above separately for each datapoint. The result of this calculation is incorporated into all relevant error bars shown in this work. We illustrate this error below using the 514 K $CH_4$ data in table S1 and figures S3 – S5. Table S1 shows the error $\Delta\delta$ obtained using equation (S14) for each datapoint in this set. The errors in pressure are obtained from the discrepancy between the pressures measured before and after the collection of the data for each volume measurement.

| Pressure (GPa) | ΔP (GPa) | δ (µm) | Δδ (µm) |
|---|---|---|---|
| 0.651 | 0.093 | 102.51 | 0.15 |
| 1.076 | 0.034 | 97.17 | 0.07 |
| 1.741 | 0.002 | 99.72 | 0.09 |
| 2.373 | 0.005 | 97.01 | 0.09 |
| 2.766 | 0.073 | 99.23 | 0.16 |
| 3.021 | 0.030 | 96.97 | 0.12 |
| 3.535 | 0.006 | 97.01 | 0.09 |
| 4.043 | 0.023 | 95.36 | 0.19 |
| 4.939 | 0.028 | 94.23 | 0.21 |

Table S1. Tabulated data for $CH_4$ at 514 K: Pressure measurements, errors in pressure measurements, measurements of δ and errors in measurements of δ.

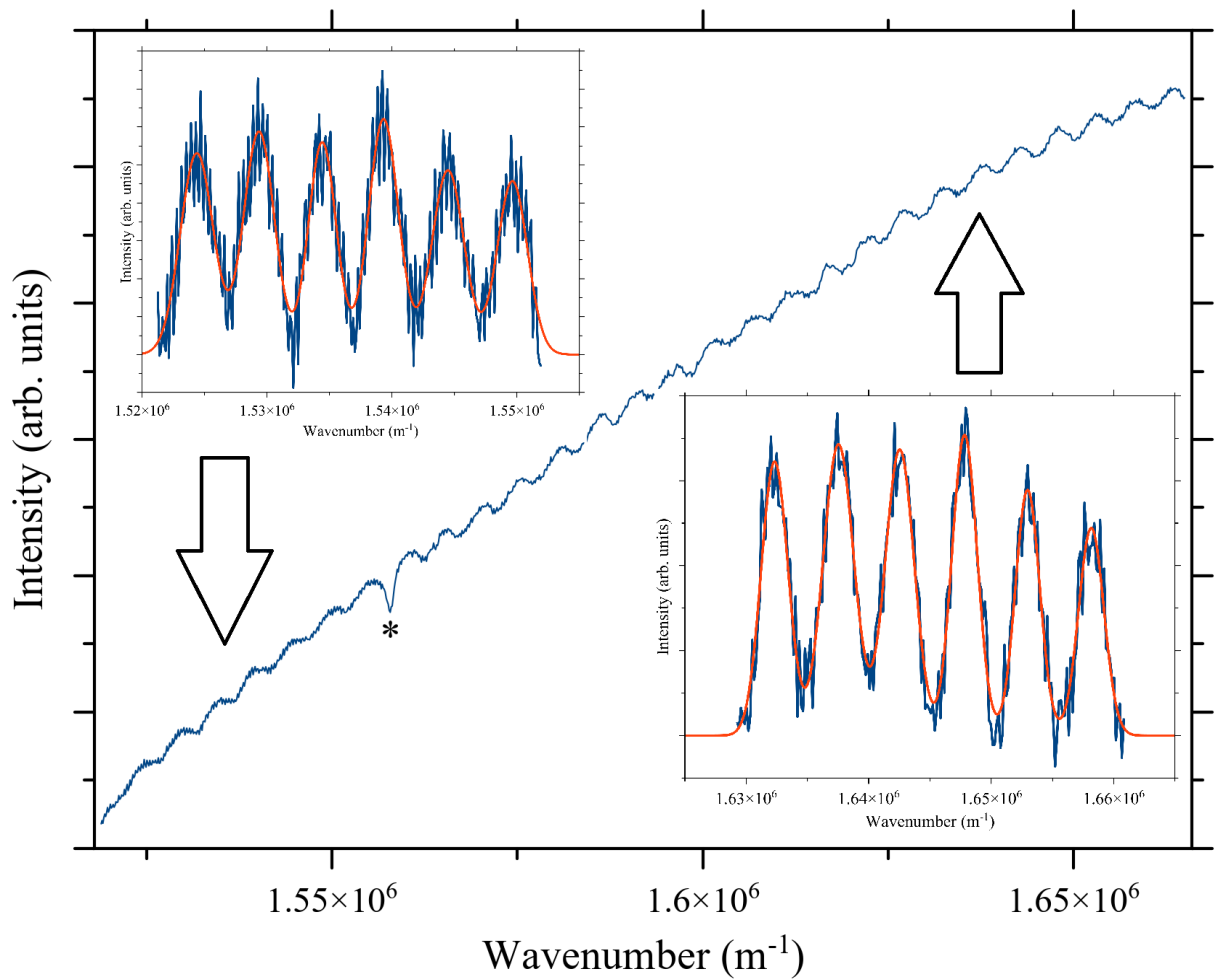

Figure S3. White light fringes from CH$_4$ sample at 514 K, 1.08 GPa. The insets show the sets of fringes which were fitted at the high and low wavenumber end of the spectrum. Here, $p = 27$. The * indicates a glitch in the CCD detector.

Figure S3 shows the white light fringe data at 1.08 GPa, 514 K. Here (and for the fringes at each other pressure) it was sometimes difficult (due to glitches on the detector) to directly count the number of fringes by visual inspection. We therefore checked the number of fringes by fitting Gaussians (after background subtraction) to ca. 5 fringes at the upper and lower wavenumber end of the spectrum. This fitting allowed us to calculate a preliminary value for δ, and hence obtain a value for $p$ to check against our visual count of the number of fringes. The Gaussian fits to the very highest and lowest wavenumber fringes in the entire spectrum were then used to calculate δ and Δδ using equations (3) and (S14). Figures S4 and S5 show the equivalent data at 3.02 GPa and 4.94 GPa.

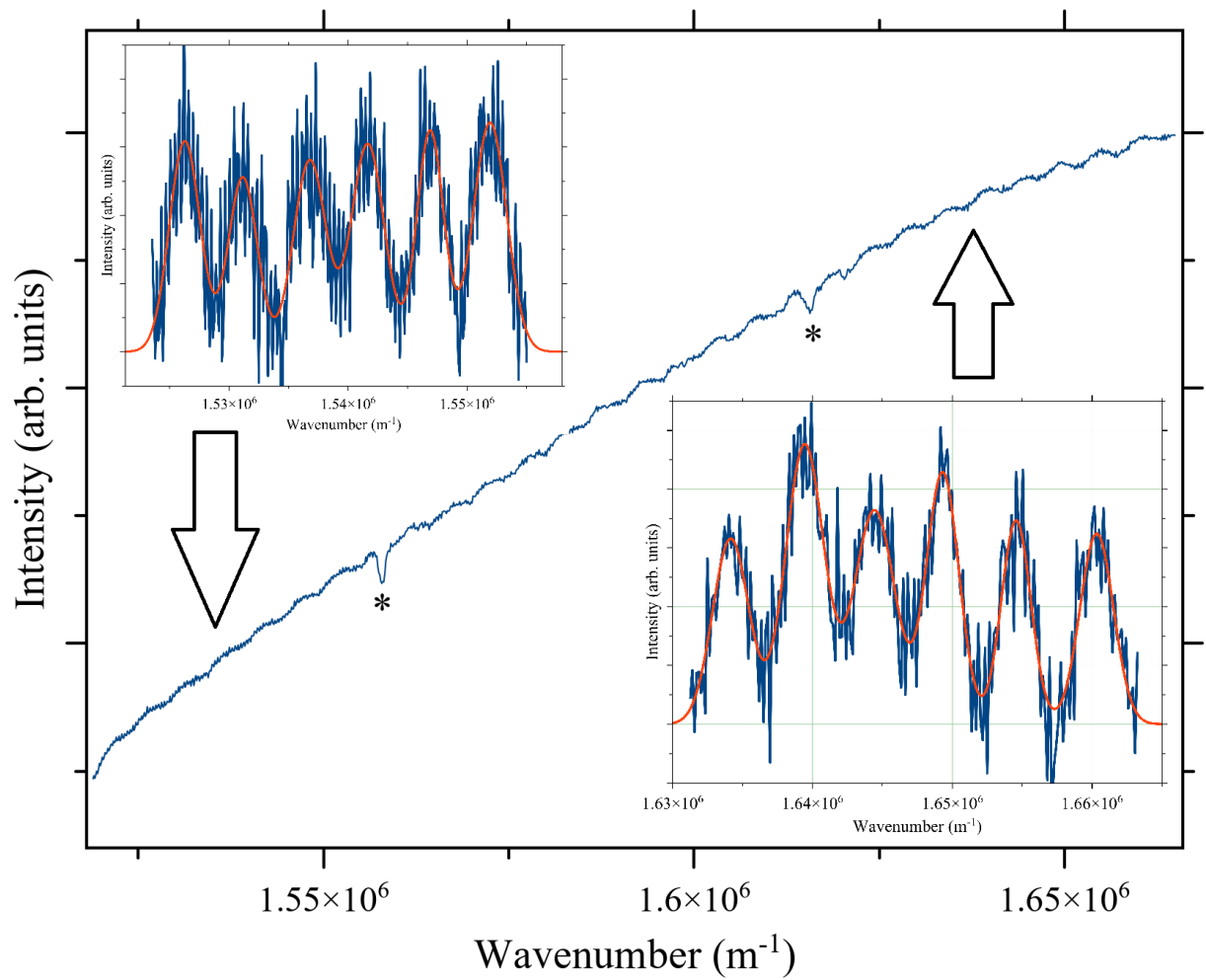

Figure S4. White light fringes from CH$_4$ sample at 514 K, 3.02 GPa. The insets show the sets of fringes which were fitted at the high and low wavenumber end of the spectrum. Here, $p = 27$. The * indicates glitches in the CCD detector.

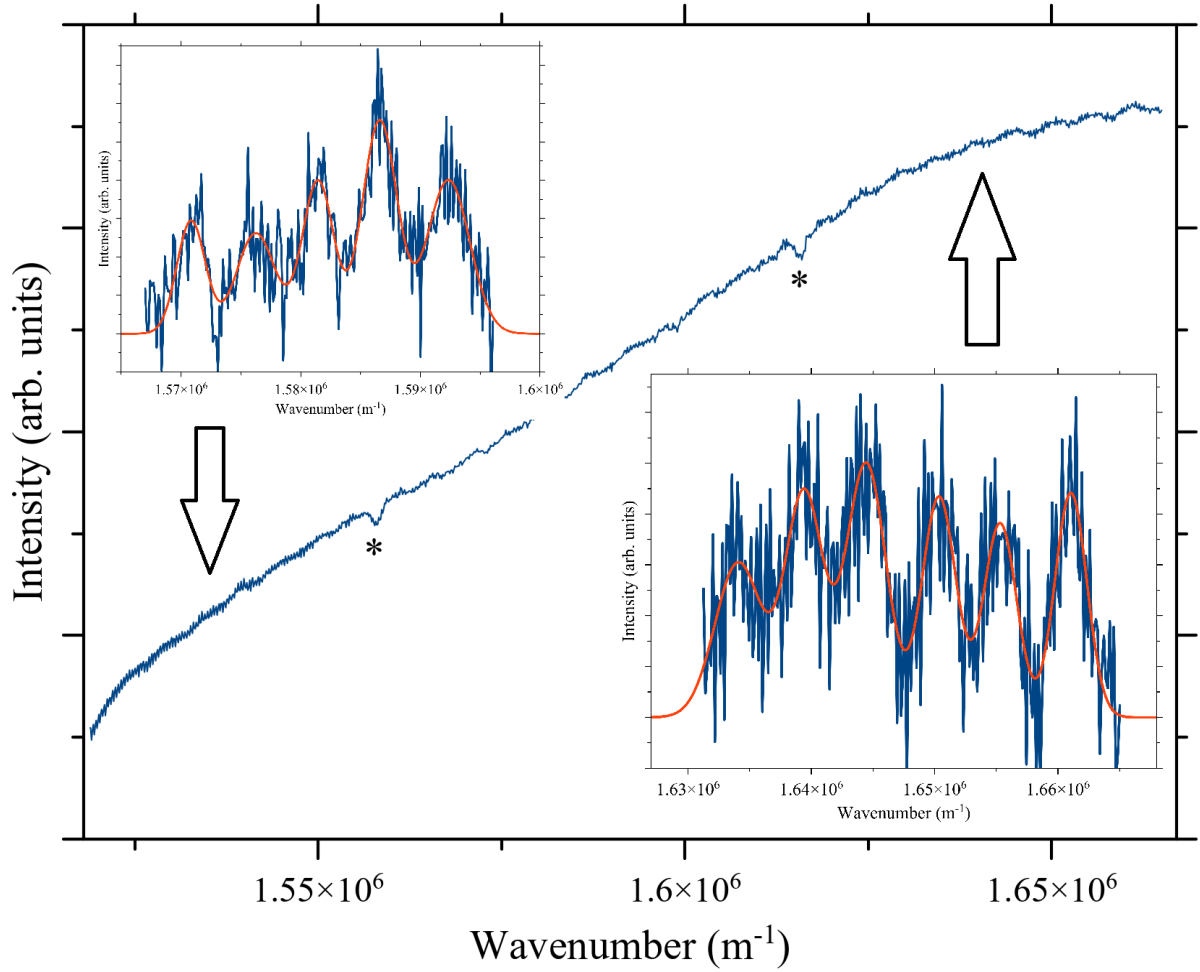

Figure S5. White light fringes from CH$_4$ sample at 514 K, 4.94 GPa. The insets show the sets of fringes which were fitted at the high and low wavenumber end of the spectrum. Here, $p = 17$. The * indicates glitches in the CCD detector.

The absolute error in $t'$ ($\Delta t'$) was estimated as ± 3 µm by conducting a number of successive measurements at the same location. The errors in $t'$ and $\delta$ propagate through to give errors in $n$ and $t$ as follows:

$$\Delta t = \frac{1}{2}\sqrt{\frac{\delta}{t'}(\Delta t')^2 + \frac{t'}{\delta}(\Delta \delta)^2}$$

$$\Delta n = \frac{1}{2}\sqrt{\frac{(\Delta \delta)^2}{t'\delta} + \frac{\delta(\Delta t')^2}{(t')^3}}$$

(S15)

Examining the equations for $\Delta t$ and $\Delta n$, we can see that since $\delta \approx t'$ and $\Delta t' \gg \Delta \delta$ in virtually all cases, the error in $t'$ is the dominant cause of error in $t$ and $n$. If we neglect the $\Delta \delta$ contribution to $\Delta n$ (but make no other approximations) we obtain:

$$\frac{\Delta n}{n} \approx \frac{1}{2}\frac{\Delta t'}{t'}$$

(S16)

The fractional error $\frac{\Delta t'}{t'}$ is significant since $t'$ (which stays smaller than $t$) varies from ca. 30 – 80 μm. Thus the fractional error $\frac{\Delta n}{n}$ varies from 2% - 5%.

Finally, the error in the cross-section area $A$ must be accounted for to obtain the error in the raw sample chamber volume $V_S$. We propose that this error is more appropriately accounted for as a fixed percentage of $A$, than as a fixed absolute value. A reasonable estimate of the uncertainty in where to fix the limits for the integration of the number of bright pixels in the images indicates an uncertainty of ± 1.5% in $A$. However, additional sources of uncertainty in this measurement are hard to quantify. In particular, the fact that the sample chamber walls may not be perfectly perpendicular to the image plane, and the fact that tiny specks of Ruby or Sm:SrB$_4$O$_7$ in the pressure chamber may reduce the bright pixel count. We therefore propose that ± 2.5% is a more reasonable estimate of the error in $A$. The amount of scatter in the plots of $A$ versus $P$ is consistent with this estimate. In this case, the overall error in the raw sample chamber volume is given by:

$$\Delta V_S = A \sqrt{(\Delta t)^2 + \left(\frac{\Delta A}{A}\right)^2 t^2}$$

(S17)

Typically, $t \approx 50 - 100 \mu m$. In this case,

$$\left(\frac{\Delta A}{A}\right)^2 t^2 \approx (\Delta t)^2$$

(S18)

Therefore, the errors in $t'$ and $A$ are both significant in causing the error $\Delta V_S$ in the raw volume data.

**Refractive index versus pressure for C$_2$H$_6$ and C$_3$H$_8$**

We conducted two experiments on C$_2$H$_6$ at ca. 380 K, in which RI data were collected on pressure increase and decrease up to 4 GPa, including one measurement in the solid state. Cross-section data were not collected and no fitting was performed on the RI data. Figure S6 shows the raw RI data as a function of pressure. We also performed an experiment on C$_3$H$_8$ at 295 K in which RI data were collected on pressure increase and decrease. Figure S6 shows the raw RI data as a function of pressure. According to the literature, the C$_3$H$_8$ fusion curve is at 3.2 GPa at ambient temperature [49]. Our own investigations indicate that the fusion curve lies at considerably lower pressure than this, perhaps as low as 1.3 GPa. This would seem to be more in line with the fusion curve above ambient temperature presented in the same reference.

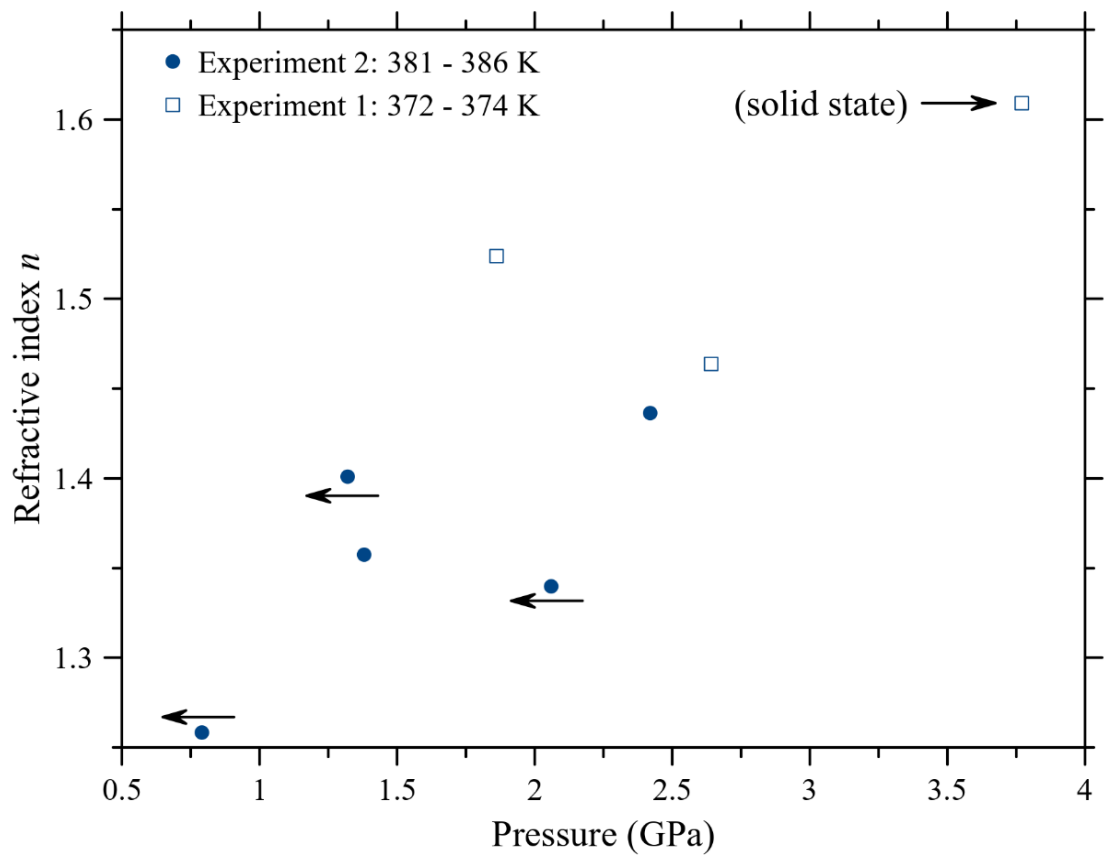

Figure S6. Raw RI data as a function of pressure for $C_2H_6$ at ca. 380 K. The arrows indicate data collected on pressure decrease.

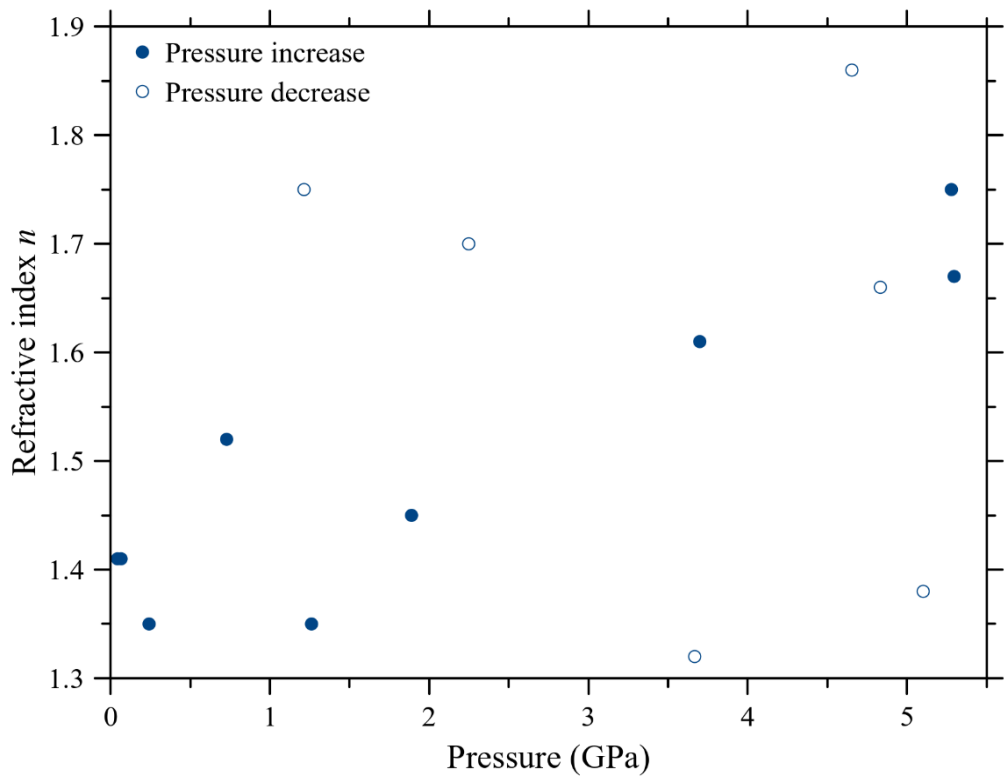

Figure S7. Raw RI data as a function of pressure for $C_3H_8$ at ca. 295 K.

**Preliminary experiments on NaCl**

In some preliminary experiments we determined the NaCl EOS using the white light interference fringes to obtain $nt$, combining this with the known RI at high pressure [15] to obtain t and our cross-section measurement to obtain $V$. These data, along with the known EOS [16], are shown in figure S8 below. Since these data were collected we significantly improved our methodology so these data should not be considered representative of the accuracy possible with our method.

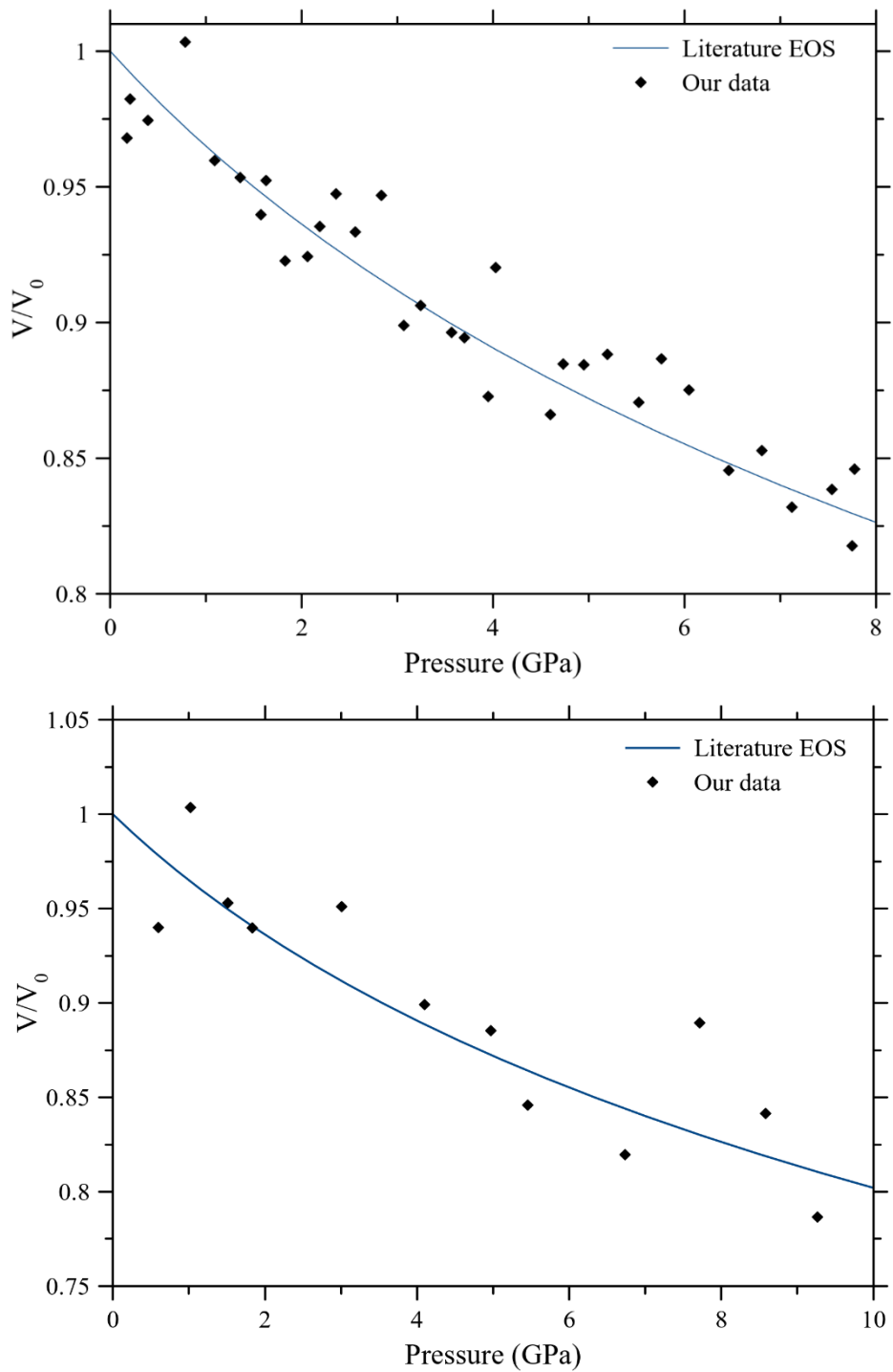

Figure S8. NaCl EOS data calculated using the white light interference fringes and known RI [15], compared to the known EOS [16]. The upper panel is data from two DAC loadings (with volumes rescaled accordingly) and the lower panel is data from a single DAC loading.

**Further experimental details**

High pressure was generated using custom-constructed piston-cylinder DACs equipped with diamonds having 450 μm or 600 μm diameter culets. For the experiments at ambient temperature the DACs utilized consisted of a short (ca. 4 cm) piston and cylinder constructed from stainless steel with pressure applied via screws or a lever arm. For the experiments at high temperature we used a DAC with a longer piston and cylinder constructed from Maraging steel, with pressure applied via a lever arm at high temperature. In both cases the DACs were gasketed using stainless steel gaskets, with sample chambers eroded using a custom-constructed electric discharge machine following pre-indentation.

High temperature was generated using a heater clamped around the cylinder of the DAC (Watlow Inc.) connected to a custom-constructed temperature controller which ensured temperature remained constant within the error margins given elsewhere.

Photoluminescence spectra of Ruby and Sm:$SrB_4O_7$ (for pressure measurement) were collected on a 1200 lines/mm conventional single grating spectrometer. Photoluminescence was excited using 405 nm and 532 nm laser diodes. In both cases the 180° backscattering geometry was used, with a 20x magnification objective (numerical aperture of 0.30). White light interference spectra (such as that shown in figure 1 of the main text) were collected on the same spectrometer, with excitation from a white LED placed behind the DAC. The dimensions of the LED and DACs limit the beam divergence to 0.04 radians (2.3°) for the experiments at ambient temperature and 0.02 radians (1.1°) for the experiments at high temperature.

The measurements of $t'$ made by recording the laser focal points were performed using a 632.8 nm HeNe laser. This laser was focussed on the sample using the same 20x / 0.30 NA objective as used for the photoluminescence spectra and white light interference patterns. The laser was operated at very low power and great care was taken to align the beam expander to produce a clean and crisply focussed beam.

KCl was loaded into the DAC by placing a small number of crystals on the end of a needle, and distilled $H_2O$ was loaded into the DAC by placing a small droplet on the end of a needle. $CH_4$, $C_2H_6$ and $C_3H_8$ were loaded into the DAC cryogenically using the same procedure as in our earlier work on $CH_4$ (e.g. ref. [33]).

A diagram of our apparatus is shown in figure S9. Additional mirrors for the purpose of directing light around the optical table have been omitted for clarity.

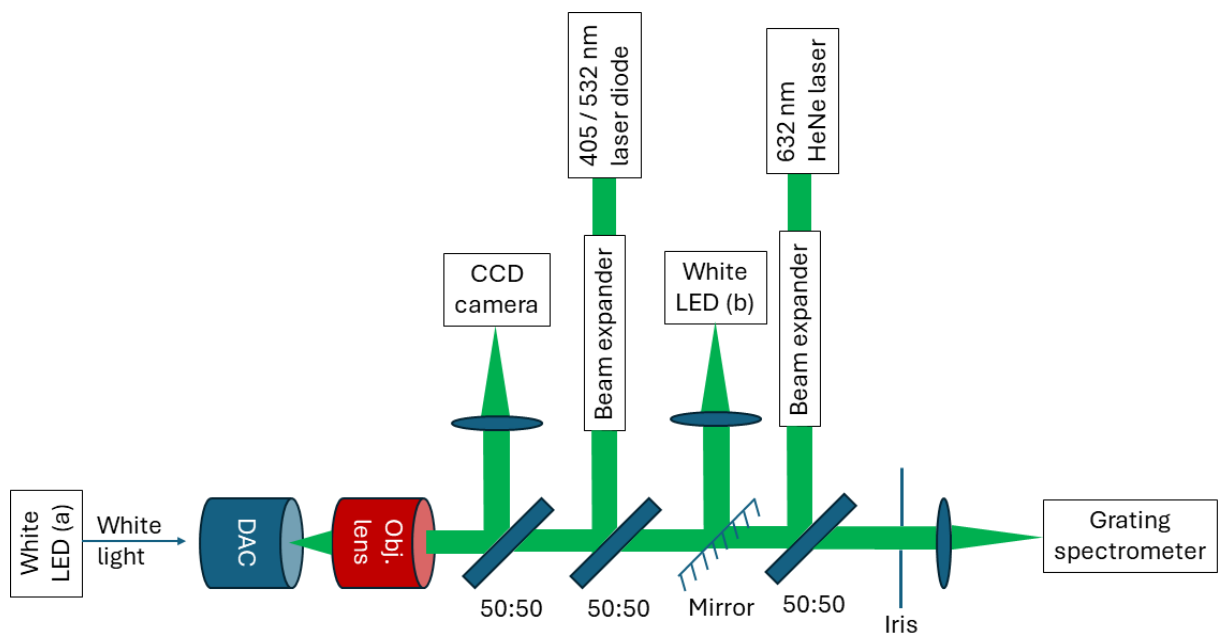

Figure S9. Apparatus diagram. 50:50 beamsplitters and mirror shown here are mounted on flip mounts so are removable from the beam. White LEDs (a) and (b) illuminate the sample with transmitted and reflected light respectively. The white light interference spectrum is collected using the transmitted light from LED (a). the CCD camera is for sample viewing and $t'$ measurements.

**Processing of CH$_4$ RI data from the literature**

The data from the literature comprising figure 7 in the main text were processed as follows:

- Hebert et al. [45], Shimizu et al. [46] and Martonchik and Orton [48] liquid state data. The pressure and RI values were digitized from the graphs in the papers. Densities were calculated from the pressure using the fundamental EOS [1] as implemented by NIST [6].
- Olson [47]. In this work the RI at the critical point was obtained. We used the critical density given by the fundamental EOS / NIST [1][6].
- Hebert et al. [45] and Shimizu et al. [46] solid phase I data. The pressure and RI values were digitized from the graphs in the papers. Densities were calculated from the pressures using our own Murnaghan fit to the X-ray diffraction data of Hazen et al. [38].
- Hebert et al. [45] solid phase A data. The pressure and RI values were digitized from the graphs in the papers. Densities were calculated from the pressures using our Murnaghan fit to the X-ray diffraction data of Nakahata et al. [50].
- Luna et al. phase II data [44]. The RI and density values are given in tabulated form in the paper, and were used without any processing or adjustment.
- Achtermann et al. gas data [42]. Although ref. [42] includes values for the densities, these are calculated from the RI assuming that the Lorentz-Lorenz law is obeyed. Instead, we therefore obtained the densities from the given $P, T$ values using the fundamental EOS / NIST [1][6] (published after ref. [42]). The data included from ref. [42] cover a wide range of temperatures, at 34 MPa (the highest pressure reached in this study). Data from this study at lower pressures were excluded from figure 7 for clarity, and to avoid overfitting the Lorentz-Lorenz law to low density data at the expense of fitting well to data at the densities covered in the present study.

Note: Data from refs. [8], [45] and [46] in the liquid state at ambient temperature above 1 GPa were excluded from figure 7 because the fundamental equation of state [1] is not backed by any experimental PV data above 1 GPa so it is not possible to calculate density from pressure without some element of extrapolation.